\documentclass{emulateapj}
\usepackage{verbatim}
\def\gsim{\mathrel{\rlap{\lower4pt\hbox{\hskip1pt$\sim$}}
    \raise1pt\hbox{$>$}}}                

\slugcomment{Accepted for publication in ApJ}

\shorttitle{FIR CO in NGC 1068}
\shortauthors{Hailey-Dunsheath et al.}

\begin{document}


\title{Herschel-PACS Observations of Far-IR CO Line Emission in NGC 1068: Highly Excited Molecular Gas in the Circumnuclear Disk\footnotemark[$\bigstar$]} \footnotetext[$\bigstar$]{Herschel is an ESA space observatory with science instruments provided by European-led Principal Investigator consortia and with important participation from NASA.}


\author{S. Hailey-Dunsheath\altaffilmark{1,2}, E. Sturm\altaffilmark{1}, J. Fischer\altaffilmark{3}, A. Sternberg\altaffilmark{4}, J. Graci{\'a}-Carpio\altaffilmark{1}, R. Davies\altaffilmark{1}, E. Gonz{\'a}lez-Alfonso\altaffilmark{5}, D. Mark\altaffilmark{4}, A. Poglitsch\altaffilmark{1}, A. Contursi\altaffilmark{1}, R. Genzel\altaffilmark{1}, D. Lutz\altaffilmark{1}, L. Tacconi\altaffilmark{1}, S. Veilleux\altaffilmark{6,7}, A. Verma\altaffilmark{8}, and J. A. de Jong\altaffilmark{1}}

\altaffiltext{1}{Max-Planck-Institut f{\"u}r extraterrestrische Physik, Postfach 1312, D-85741 Garching, Germany.}
\altaffiltext{2}{Current Address: California Institute of Technology, Mail Code 301-17, 1200 E. California Blvd., Pasadena, CA 91125, USA; shd@astro.caltech.edu.}
\altaffiltext{3}{Naval Research Laboratory, Remote Sensing Division, 4555 Overlook Ave SW, Washington, DC 20375, USA.}
\altaffiltext{4}{Tel Aviv University, Sackler School of Physics \& Astronomy, Ramat Aviv 69978, Israel.}
\altaffiltext{5}{Departamento de Fisica, Universidad de Alcal{\'a} de Henares, 28871 Alcal{\'a} de Henares, Madrid, Spain.}
\altaffiltext{6}{Department of Astronomy, University of Maryland, College Park, MD 20742, USA.}
\altaffiltext{7}{Astroparticle Physics Laboratory, NASA Goddard Space Flight Center, Code 661, Greenbelt, MD 20771 USA}
\altaffiltext{8}{Department of Astrophysics, Oxford University, Oxford OX1 3RH, UK.}


\begin{abstract}
We report the detection of far-IR CO rotational emission from the prototypical Seyfert 2 galaxy NGC 1068. Using Herschel-PACS, we have detected 11 transitions in the $J_\mathrm{upper}=14-30$ ($E_\mathrm{upper}/k_B = 580-2565$ K) range, all of which are consistent with arising from within the central 10$\arcsec$ (700 pc). The detected transitions are modeled as arising from 2 different components: a moderate excitation (ME) component close to the galaxy systemic velocity, and a high excitation (HE) component that is blueshifted by $\sim$$80$ km\,s$^{-1}$. We employ a large velocity gradient (LVG) model and derive ${n_\mathrm{H}}_2\sim10^{5.6}$ cm$^{-3}$, $T_\mathrm{kin}\sim170$ K, and ${M_H}_2\sim10^{6.7}$ $M_\odot$ for the ME component, and ${n_\mathrm{H}}_2\sim10^{6.4}$ cm$^{-3}$, $T_\mathrm{kin}\sim570$ K, and ${M_H}_2\sim10^{5.6}$ $M_\odot$ for the HE component, although for both components the uncertainties in the density and mass are $\pm(0.6-0.9)$ dex. Both components arise from denser and possibly warmer gas than traced by low-$J$ CO transitions, and the ME component likely makes a significant contribution to the mass budget in the nuclear region. We compare the CO line profiles with those of other molecular tracers observed at higher spatial and spectral resolution, and find that the ME transitions are consistent with these lines arising in the $\sim$$200$ pc diameter ring of material traced by H$_2$ 1-0 S(1) observations. The blueshift of the HE lines may also be consistent with the bluest regions of this H$_2$ ring, but a better kinematic match is found with a clump of infalling gas $\sim$$40$ pc north of the AGN. We consider potential heating mechanisms, and conclude that X-ray or shock heating of both components is viable, while far-UV heating is unlikely. We discuss the prospects of placing the HE component near the AGN, and conclude that while the moderate thermal pressure precludes an association with the $\sim$$1$ pc radius H$_2$O maser disk, the HE component could potentially be located only a few parsecs more distant from the AGN, and might then provide the $N_H\sim10^{25}$ cm$^{-2}$ column obscuring the nuclear hard X-rays. Finally, we also report sensitive upper limits extending up to $J_\mathrm{upper}=50$, which place constraints on a previous model prediction for the CO emission from the X-ray obscuring torus.
\end{abstract}





\section{Introduction}

The excited molecular gas in the centers of Seyfert galaxies offers a sensitive probe of the nature of active galactic nucleus (AGN) feedback on the surrounding interstellar medium (ISM). Observational studies of the most highly excited material in Seyfert nuclei have typically used the H$_2$ rotational~\citep{Lutz2000all,Rigopoulou2002,Roussel2007} and the well-studied ro-vibrational~\citep[e.g.,][]{Thompson1978,Mouri1994,Maloney1997,Davies2005,Rodriguez_Ardila2005} transitions. The pure H$_2$ rotational lines ($E_\mathrm{upper}/k_\mathrm{B}\gsim500$ K) are easily thermalized at moderate (${n_\mathrm{H}}_2\gsim10^3$ cm$^{-3}$) densities, while the ro-vibrational lines ($E_\mathrm{upper}/k_\mathrm{B}\gsim7000$ K) may be excited through collisions in dense (${n_\mathrm{H}}_2\gsim10^5$ cm$^{-3}$) gas or through UV fluorescence~\citep{Sternberg1989}. Observations of these tracers in Seyferts have identified a number of potentially important excitation mechanisms, including X-rays from the AGN~\citep{Maloney1997,Rodriguez_Ardila2005}, shocks associated with supernova remnants, radio jets, and gravitational instabilities~\citep{Roussel2007}, and stellar far-UV (FUV) radiation~\citep{Davies2005}, with no clear consensus on a single dominant excitation source. The far-IR (FIR) CO rotational transitions (CO[$J_\mathrm{upper}\rightarrow J_\mathrm{upper}-1$], with $J_\mathrm{upper}\approx13-50$) arise from states $500-7,000$ K above ground and have critical densities of $\sim10^6-10^8$ cm$^{-3}$, and complement the H$_2$ transitions for studies of warm and dense material. Compared with H$_2$, the FIR CO lines trace similar energy levels, but have higher critical densities, and are less sensitive to extinction. Additionally, the smaller energy gaps between levels leads to a finer sampling of density-temperature phase space. These lines have been proposed as potential tools for studying the obscuring medium of type 2 AGN~\citep{Krolik1989}, determining the energy budgets of composite starburst/AGN systems~\citep{Meijerink2007}, and identifying accreting black holes in the early universe~\citep{Spaans2008,Schleicher2010alma}, but previous facilities were unable to detect this line emission from extragalactic sources. Here we take advantage of the superb sensitivity of Herschel-PACS to conduct the first extragalactic study of FIR CO emission, from the prototypical Seyfert 2 galaxy NGC 1068.

NGC 1068 is one of the brightest and best studied Seyfert 2 galaxies. The paradigm of an optically and geometrically thick molecular torus accounting for the Seyfert type 1 and 2 dichotomy followed the detection of scattered broad line emission from this source~\citep{Miller1983scatter}, and NGC 1068 has been at the center of subsequent studies of the ISM in Seyfert nuclei. The molecular gas in the central 1$\arcmin$ of NGC 1068 has been well studied, and here we review some of the key results. Interferometric observations of CO(1-0) have identified a pair of $\approx$$15\arcsec$ ($\approx$$1.1$ kpc) radius spiral arms~\citep{Planesas1991,Helfer1995,Schinnerer2000}, which may be modeled as forming in response to a $\approx$$17$ kpc bar~\citep{Schinnerer2000}. These arms are bright in Br$\gamma$~\citep{Davies1998}, PAH emission~\citep{LeFloch2001}, and submillimeter continuum~\citep{Papadopoulos1999SCUBA}, and contain most of the star formation in the central region. Centered on the AGN is the $\sim$$5\arcsec$ ($\sim$$350$ pc) circumnuclear disk (CND), which is visible in CO and H$_2$ 1-0 S(1), but becomes particularly prominent in images of HCN~\citep{Tacconi1994} and other high density tracers~\citep{GarciaBurillo2010}. Strong emission in CO(4-3) and HCN(1-0) indicate the gas in the CND is both warm and dense~\citep[][although see~\citet{,Krips2011} for a lower density model]{Tacconi1994,Sternberg1994,Israel20091068}. The high abundances of HCN, CN, H$_3$O$^+$, and other molecules in the CND suggest an X-ray$-$driven chemistry~\citep{Usero2004,GarciaBurillo2010,Aalto2011}, and X-ray heating has also been invoked to explain the strong H$_2$ 1-0 S(1) and [Fe II] emission~\citep{Rotaciuc1991,Maloney1997,Galliano2002}. At $\sim$$0.3\arcsec$ resolution the H$_2$ 1-0 S(1) observations resolve the CND into a $\sim$$1\arcsec$ ring-like structure~\citep{Galliano2002}, while the line spectral profiles show evidence for rotation, expansion, and more complex kinematics~\citep{Galliano2002,Galliano2003,Davies2008conference}. Shocks following this non-circular motion, and possibly associated with jet-ISM interactions, may also be important in heating the molecular CND~\citep{Krips2011}. At $\sim$$0.1\arcsec$ resolution the H$_2$ 1-0 S(1) images reveal two clumps of infalling molecular material at $\sim$$0.1-0.4\arcsec$ scales that likely play an important role in both fueling and obscuring the AGN, with an estimated infall rate of $\sim$$15$ $M_\odot$\,yr$^{-1}$ to within a few parsecs of the nucleus~\citep{MuellerSanchez2009}. Finally, milli-arcsec resolution radio observations identify a series of H$_2$O maser spots that trace out the inner surface of a $\sim$$0.65$ pc radius molecular disk, centered on the AGN~\citep[][and references therein]{Gallimore2004}.

Hard X-ray observations of NGC 1068 indicate large obscuration to the nucleus~\citep{Iwasawa1997, Matt19971068, Colbert2002} by an intervening medium with a column density possibly exceeding $N_\mathrm{H}>10^{25}$ cm$^{-2}$~\citep{Matt19971068}. Interferometric mid-IR observations of NGC 1068 identify a parsec scale structure of hot dust that, along with the H$_2$O maser disk, may represent the dusty molecular torus responsible for the X-ray obscuration~\citep{Jaffe2004,Raban2009}. However, other investigators have found evidence that at least some of the nuclear obscuration occurs on few to ten parsec scales from the AGN~\citep{Cameron1993,Hoenig2006,MuellerSanchez2009}.

The organization of this paper is as follows. In section~\ref{observationssection} we describe the Herschel-PACS observations of the FIR CO lines in NGC 1068. In section~\ref{resultssection} we analyze the gas excitation and estimate physical parameters, and in section~\ref{comparisonsection} we compare the physical parameters and line profiles with those of other molecular gas tracers. In sections~\ref{meheating} and~\ref{heheating} we discuss potential heating mechanisms. In section~\ref{nondetections} we discuss our detections and upper limits in the context of the molecular ISM within a few parsecs of the AGN, and in section~\ref{conclusions} we summarize our findings. Throughout this paper we adopt a distance to NGC 1068 of 14.4 Mpc~\citep{Bland_Hawthorn1997}, and a systemic velocity $V_\mathrm{LSR} = 1125$ km\,s$^{-1}$. 

\section{Observations} \label{observationssection}

\subsection{Data Acquisition and Reduction}

The observations were made with the Photodetector Array Camera and Spectrometer~\citep[PACS;][]{Poglitsch2010} on board the \textit{Herschel} Space Observatory~\citep{Pilbratt2010}. Most of the data presented here was obtained as part of the SHINING guaranteed time key program. The SHINING observations consisted of ten high resolution range scans concatenated to cover the $52-98$ $\mu$m and $104-196$ $\mu$m ranges, as well as deeper integrations of CO(17-16), CO(24-23), and CO(40-39). These latter observations targeted CO transitions falling in relatively clean spectral regions, and were conducted to provide a coarse but sensitive sampling of the CO SED over the full FIR range. The SHINING data yielded detections of most transitions at $J_\mathrm{upper} \le 24$, and we obtained follow-up observations of CO(28-27) and CO(30-29) in an open time project to extend our CO SED measurements to higher-$J$. These observations amounted to a total of 13.7 hours of integration time. The data reduction was done using the standard PACS reduction and calibration pipeline (ipipe) included in HIPE 5.0 975. However, for the final calibration we normalized the spectra to the telescope flux and recalibrated it with a reference telescope spectrum obtained from dedicated Neptune continuum observations. With this approach we estimate an absolute flux calibraton accuracy of 30\%.

\subsection{Line Flux Estimation}

\begin{figure}
\epsscale{1.1}
\plotone{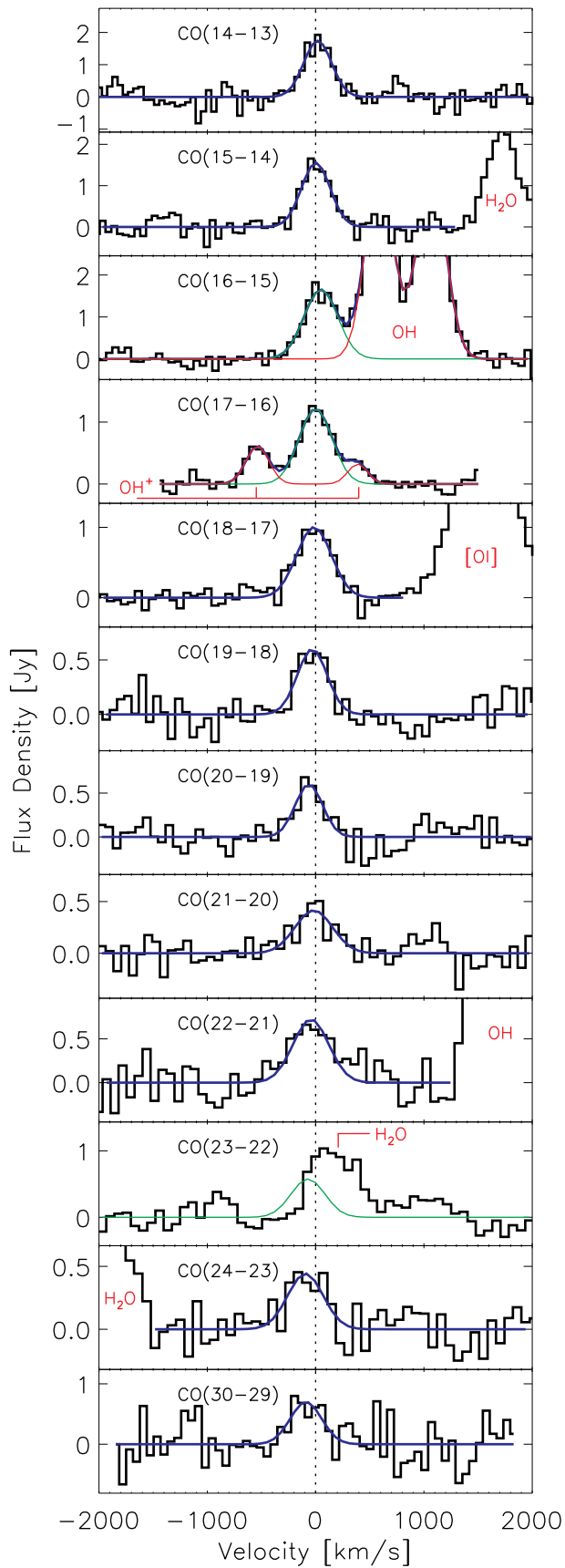}
\caption{Continuum-subtracted spectra of the $J_\mathrm{upper}=14-24$ and $J_\mathrm{upper}=30$ CO lines. The blue curves show the line fits, and features other than CO are labeled in red. For CO(16-15) and CO(17-16) the green and red curves show the decomposition of the fit into CO and other lines, respectively. All lines are detected with the exception of CO(23-22), which is blended with a strong H$_2$O line $209$ km\,s$^{-1}$ to the red. Here the overplotted green curve is an average of the CO(22-21) and CO(24-23) profiles as a reference.\label{spectra}}
\end{figure}

The PACS spectrometer performs integral field spectroscopy over a $47\arcsec\times47\arcsec$ FOV, resolved into a $5\times5$ array of $9.4\arcsec$ spatial pixels (spaxels). The spectrometer resolving power varies from $R=1000-3000$ for the 1$^\mathrm{st}$ and 2$^\mathrm{nd}$ order observations utilized here. In Figure~\ref{spectra} we show the spectra from the central spaxel centered on 11 of the 12 CO transitions falling in the $104-196$ $\mu$m range. The CO(25-24) line at $\lambda_\mathrm{rest}=104.44$ $\mu$m lies in a noisy region at the edge of this range, and is not included. Most of these lines are strong in the central spaxel, but little flux is detected outside of the central spaxel, as expected for an unresolved source ($\theta_\mathrm{source} < 9.4\arcsec$). All fluxes and upper limits presented here were therefore extracted from the central spaxel, and referenced to a point source by dividing by the recommended point source correction factors\footnote{see also http://herschel.esac.esa.int/twiki/pub/Public/PacsCalibrationWeb/PacsSpectroscopyPerformanceAndCalibration\_v2\_4.pdf}~\citep{Poglitsch2010}. 

The CO line fluxes were measured by fitting the spectra with a Gaussian profile plus a baseline. In most cases a linear baseline was adopted and the three parameters defining the Gaussian were allowed to vary freely, but some lines required a modified approach. A broad feature underneath the CO(15-14) line is present in the raw data, likely due to an imperfect subtraction of the telescope background, and we remove this feature using a higher order baseline fit. The integrated CO(15-14) flux and the residual line profile shown in Figure~\ref{spectra} are consistent with those of adjacent transitions, and we estimate that the flux uncertainty introduced by this baseline feature is less than the assumed 30\% absolute flux calibration error. The CO(16-15) line is blended with the 163 $\mu$m OH doublet, and CO(17-16) with a pair of flanking OH$^+$ lines. In both cases we estimate the CO flux by simultaneously fitting all features. An unconstrained Gaussian fit to the relatively low S/N CO(22-21) line yields a much broader profile than for any other transition, so here we fix the width of the CO profile to the typical value derived from other line fits (corresponding to an intrinsic FWHM of $250$ km\,s$^{-1}$; see Figures~\ref{coSEDLVG} and~\ref{velcentroid}, and the discussion in section~\ref{pacslinekinematics}). CO(23-22) is blended with a strong H$_2$O line with a rest wavelength 209 km\,s$^{-1}$ to the red. If the combined feature were attributed solely to H$_2$O it would be both broader and more blueshifted than any of the other 6 H$_2$O lines detected in the PACS scans, and we interpret this as evidence for significant contamination by CO(23-22). A comparison with the average of the CO(22-21) and CO(24-23) profiles suggests the CO(23-22) line is weaker and/or more redshifted than these lines (Figure~\ref{spectra}). However, due to the uncertainties involved in deconvolving the CO and H$_2$O lines, we simply exclude CO(23-22) from our analysis. All other transitions from CO(14-13) through CO(24-23) are well detected (Table~\ref{fluxtable}).
%
 
The $52-98$ $\mu$m range includes the CO(27-26) through CO(50-49) transitions. None of these lines are detected in the full range scans or the targeted CO(40-39) observation obtained with SHINING, but our follow-up open time program yielded a detection of CO(30-29) (Figure~\ref{spectra}). The flux for this line was estimated using the same fitting procedure as described above for the $J_\mathrm{upper} \le 24$ lines (Table~\ref{fluxtable}). We estimate upper limits for the nondetected transitions by first binning the data to $600$ km\,s$^{-1}$ bins, and then estimating the $3\sigma$ noise levels (Table~\ref{fluxtable}). 

We detect no emission from $^{13}$CO. The $^{13}$CO(14-13) transition at $\lambda_\mathrm{rest}=194.55$ $\mu$m lies at the noisy edge of a scan, while the $J_\mathrm{upper} \gsim 21$ transitions at $\lambda \ge 104$ $\mu$m are blended with $^{12}$CO lines. For $^{13}$CO(15-14) through $^{13}$CO(20-19) we estimate $3\sigma$ upper limits of $(2-4)\times10^{-17}$ W\,m$^{-2}$, for a $600$ km\,s$^{-1}$ bin size. Our most stringent lower limit on the $^{12}$CO/$^{13}$CO flux ratio comes from the $J_\mathrm{upper}=16$ transition, for which we estimate $^{12}$CO(16-15)/$^{13}$CO(16-15) $\gsim2.6$. Assuming the $^{12}$CO and $^{13}$CO SEDs are similar, this suggests we can exclude significant contamination of the detected $^{12}$CO lines at $J_\mathrm{upper} \gsim 21$ from $^{13}$CO.

\renewcommand{\thefootnote}{\alph{footnote}}
\begin{table}
\begin{center}
\caption{PACS CO Line Observations\label{fluxtable}}
\begin{tabular}{ccccccccccccrrrrr}
\tableline\tableline
Line	   & $\lambda_\mathrm{rest}$  & Flux\tablenotemark{a}     & $V_0$\tablenotemark{b}  & FWHM                    \\
           & [$\mu$m]                 & [$10^{-17}$ W\,m$^{-2}$]  & [km\,s$^{-1}$]          & [km\,s$^{-1}$]          \\
\tableline
CO(14-13)  & 186.00                   & $7.2 \pm 2.3$             & $19  \pm 23$            & $305 \pm 32$            \\
CO(15-14)  & 173.63                   & $6.4 \pm 2.2$             & $2   \pm 25$            & $313 \pm 37$            \\
CO(16-15)  & 162.81                   & $8.1 \pm 2.5$             & $49  \pm 25$            & $369 \pm 28$            \\
CO(17-16)  & 153.27                   & $5.8 \pm 1.8$             & $0   \pm 26$            & $359 \pm 27$            \\
CO(18-17)  & 144.78                   & $5.1 \pm 1.6$             & $-15 \pm 28$            & $379 \pm 31$            \\
CO(19-18)  & 137.20                   & $2.6 \pm 0.9$             & $-31 \pm 35$            & $324 \pm 55$            \\
CO(20-19)  & 130.37                   & $2.5 \pm 0.9$             & $-62 \pm 36$            & $297 \pm 55$            \\
CO(21-20)  & 124.19                   & $2.4 \pm 0.9$             & $-17 \pm 46$            & $407 \pm 87$            \\
CO(22-21)  & 118.58                   & $4.0 \pm 1.4$             & $-44 \pm 43$            & $387$\tablenotemark{c}  \\
CO(23-22)  & 113.46                   & blended                   &                         &                         \\
CO(24-23)  & 108.76                   & $2.6 \pm 1.0$             & $-94 \pm 49$            & $385 \pm 95$            \\
CO(25-24)  & 104.44                   & $<11.2      $             &                         &                         \\
CO(26-25)  & 100.46                   & n/a                       &                         &                         \\
CO(27-26)  & 96.77                    & $<5.8       $             &                         &                         \\
CO(28-27)  & 93.35                    & $<4.6       $             &                         &                         \\
CO(29-28)  & 90.16                    & $<9.3       $             &                         &                         \\
CO(30-29)  & 87.19                    & $4.2 \pm 1.9$             & $-89 \pm 50$            & $341 \pm 117$            \\
CO(31-30)  & 84.41                    & blended                   &                         &                         \\
CO(32-31)  & 81.81                    & $<7.4       $             &                         &                         \\
CO(33-32)  & 79.36                    & $<9.5       $             &                         &                         \\
CO(34-33)  & 77.06                    & $<5.8       $             &                         &                         \\
CO(35-34)  & 74.89                    & $<6.2       $             &                         &                         \\
CO(36-35)  & 72.84                    & $<8.3       $             &                         &                         \\
CO(37-36)  & 70.91                    & $<7.8       $             &                         &                         \\
CO(38-37)  & 69.07                    & $<10.1      $             &                         &                         \\
CO(39-38)  & 67.34                    & $<19.1      $             &                         &                         \\
CO(40-39)  & 65.69                    & $<22.3      $             &                         &                         \\
CO(41-40)  & 64.12                    & $<13.2      $             &                         &                         \\
CO(42-41)  & 62.62                    & $<21.6      $             &                         &                         \\
CO(43-42)  & 61.20                    & $<16.8      $             &                         &                         \\
CO(44-43)  & 59.84                    & $<10.8      $             &                         &                         \\
CO(45-44)  & 58.55                    & $<14.8      $             &                         &                         \\
CO(46-45)  & 57.31                    & blended                   &                         &                         \\
CO(47-46)  & 56.12                    & $<14.7      $             &                         &                         \\
CO(48-47)  & 54.99                    & $<28.3      $             &                         &                         \\
CO(49-48)  & 53.90                    & $<31.7      $             &                         &                         \\
CO(50-49)  & 52.85                    & $<45.2      $             &                         &                         \\
\tableline
\end{tabular}
\tablenotetext{1}{Total uncertainties combine a 30\% calibration error with statistical errors in line fits. Upper limits are $3\sigma$, and refer to the flux density integrated over a $600$ km\,s$^{-1}$ bin. Some lines are blended with a strong feature, and it is not possible to obtain a flux measurement or useful upper limit. CO(26-25) was not covered in the PACS scans.}
\tablenotetext{2}{Relative to $V_\mathrm{LSR}=1125$ km\,s$^{-1}$. Total uncertainties combine a spectral calibration accuracy of 10\% of the spectral resolution ($20-30$ km\,s$^{-1}$) with statistical errors in line fits.}
\tablenotetext{3}{Fixed to an intrinsic FWHM of 250 km\,s$^{-1}$ (see text).}
\end{center}
\end{table}
\renewcommand{\thefootnote}{\arabic{footnote}}

\section{Excitation Analysis} \label{resultssection}
\subsection{Evidence for Two Components} \label{twocomponents}

\begin{figure}
\epsscale{1.1}
\plotone{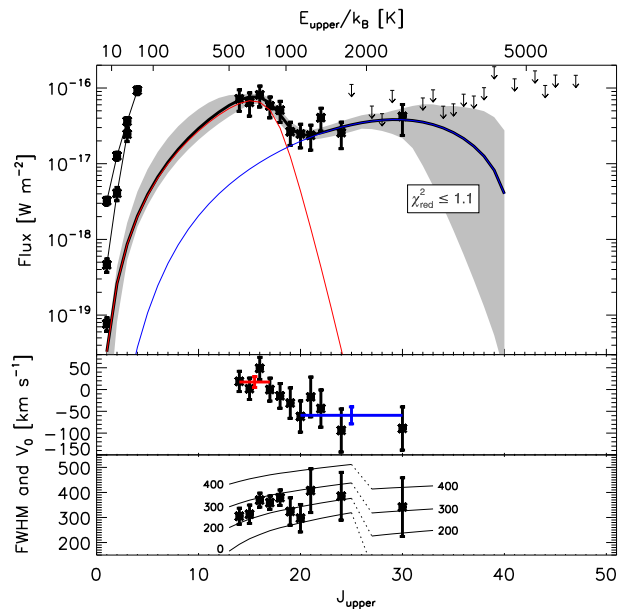}
\caption{\textit{Top:} FIR CO line fluxes and upper limits measured here, along with lower-$J$ lines from the literature. The brightest set of $J_\mathrm{upper}=1-4$ line fluxes are measured in $11\arcsec - 21\arcsec$ beams that contain a mixture of the CND and the more extended emission~\citep{Israel20091068}. The second set of $J_\mathrm{upper}=1-3$ points are interferometric measurements integrated over the central $4\arcsec$ from~\citet{Krips2011}, and the fainter CO(1-0) flux is the same from~\citet{Schinnerer2000}. The shaded region shows the range of good-fitting LVG models ($\chi^2_\mathrm{red} = \chi^2\mathrm{/dof}\le1.1$; section~\ref{goodfits}), and the solid black curve is the single best fitting model, with the red and blue curves showing the individual contributions from the ME and HE components. \textit{Middle:} Central velocities of the FIR CO lines, with average values of the ME and HE line centers indicated with horizontal red and blue lines. \textit{Bottom:} Measured FWHM of the FIR CO lines, with tracks indicating the expected line widths (estimated by adding the intrinsic line widths and the spectrometer resolution in quadrature) for sources with intrinsic widths of 0, 100, 200, and 400 km\,s$^{-1}$.\label{coSEDLVG}}
\end{figure}

In the top panel of Figure~\ref{coSEDLVG} we show the line fluxes and upper limits measured here, along with lower-$J$ measurements obtained from the literature. The middle and bottom panels show the central velocities and widths obtained from the Gaussian fitting. The inflection point seen in the FIR CO line SED at $J_\mathrm{upper}\approx19$ suggests the presence of multiple components, as does the shift in central velocities between the lowest- and highest-$J$ transitions. For simplicity we assume the FIR CO lines are produced by 2 discrete components: a moderate excitation (ME) component near the systematic velocity, and a blueshifted high excitation (HE) component. Our excitation analysis described below indicates that the $J_\mathrm{upper}\le17$ and $J_\mathrm{upper}\ge20$ transitions are dominated by the ME and HE components, respectively. Separately averaging the central velocities of these two sets of lines gives $V_\mathrm{ME}=17\pm12$ km\,s$^{-1}$ and $V_\mathrm{HE}=-59\pm20$ km\,s$^{-1}$ (Figure~\ref{coSEDLVG}), with a difference of $V_\mathrm{ME} - V_\mathrm{HE} = 76\pm23$ km\,s$^{-1}$.

\subsection{LVG Modeling} \label{lvgmodel}

To quantitatively analyze the FIR CO line SED we employ a large velocity gradient (LVG) model. We use the LVG calculation described in~\citet{HaileyDunsheath2008}, with updated CO-H$_2$ collisional coefficients from~\citet{Yang2010}, and a thermalized H$_2$ ortho/para ratio. In this model the shape of the CO line SED is determined by the gas density (${n_\mathrm{H}}_2$), kinetic temperature ($T_\mathrm{kin}$), and CO abundance per velocity gradient ([CO/H$_2$]/($dv/dr$)). The source is assumed to consist of a number of unresolved clouds, and the absolute line luminosities scale with the total CO mass ($M_\mathrm{CO}$). In the following analysis we use a CO abundance of [CO/H$_2$] $ = 10^{-4}$ to reparameterize [CO/H$_2$]/($dv/dr$) as $dv/dr$, and $M_\mathrm{CO}$ as the H$_2$ mass (${M_H}_2$). This results in an 8 parameter model, with 4 parameters for each of the ME and HE components. In section~\ref{reducedxco} we discuss the effects of varying the CO abundance.

\subsubsection{Background Radiation} \label{background}

The CO excitation may be affected by the background radiation, and we must therefore estimate the local radiation density. At millimeter wavelengths the background is dominated by the cosmic microwave background (CMB)~\citep{Kamenetzky2011}, but the FIR background arises from within the galaxy. The PACS integral field spectra provide a continuum map of the central $47\arcsec\times47\arcsec$ with $9.4\arcsec$ pixels, and demonstrate that the continuum emission in the central pixel is dominated by sources within the central $\approx$$10\arcsec$. The flux density in the central pixel may be modeled as an optically thin modified blackbody with $\beta=1.5$, $T_\mathrm{dust}=48$ K, and normalized to $F_\nu(\lambda = 100$ $\mu$m$) = 49$ Jy, and we adopt this as an estimate of the continuum brightness of the $\sim$$5\arcsec$ CND. If the flux in the central spaxel is indeed due solely to the CND than this is a moderate (by a factor of $\lesssim2$) underestimate, while emission from within the central $\approx$$10\arcsec$ but outside of the CND may also be contributing. The measured continuum is in reasonable agreement with previous calculations and observations. For comparison, this measured $F_\nu$ is a factor of $1\rightarrow4$ times larger in the $\lambda=50\rightarrow200$ $\mu$m range than obtained from the radiative transfer modeling in~\citet{Spinoglio2005}. Additionally, extrapolating our modeled SED to $\lambda=450$ $\mu$m yields $F_\nu(\lambda = 450$ $\mu$m$) = 1.2$ Jy, comparable to the peak value of $\sim$$1.5$ Jy\,beam$^{-1}$ measured in a $\sim$$9\arcsec$ beam by~\cite{Papadopoulos1999SCUBA}. The 60 $\mu$m/100 $\mu$m ratio in this model is 1.27, and the FIR flux\footnote{$F_\mathrm{FIR}=1.26\times10^{-14}\,[2.58f_\mathrm{60}/\mathrm{Jy}+f_\mathrm{100}/\mathrm{Jy}]$ W\,m$^{-2}$} is $2.7\times10^{-12}$ W\,m$^{-2}$, giving $L_\mathrm{FIR}=4\pi\mathrm{d}^2F_\mathrm{FIR}=1.7\times10^{10}$ $L_\odot$.

We include the effects of background radiation on the equations of statistical equilibrium following~\citet{Poelman2005}. The important parameter in this approach is the mean specific intensity of the external radiation field at the cloud surface, which we define as $J_{\nu\mathrm{,ext}}$. We estimate $J_{\nu\mathrm{,ext}}$ using a simple geometrical model in which the gas clouds are uniformly distributed in a sphere with an observed angular size $\Omega$, and are evenly mixed with the FIR-emitting dust grains. For optically thin continuum, the mean value of $J_{\nu\mathrm{,ext}}$ is then related to the observed continuum flux density $F_{\nu\mathrm{,obs}}$ as 

\begin{equation}
J_{\nu\mathrm{,ext}} = I_{\nu\mathrm{,CB}} + \frac{9}{16}\frac{F_{\nu\mathrm{,obs}}}{\Omega},
\label{backgroundfield}
\end{equation}

\noindent where $I_{\nu\mathrm{,CB}}$ is the sum of the CMB and cosmic IR background (CIB). We take $\Omega$ to correspond to a circular diameter of $4\arcsec$, approximately matched to the size of the CND (see section~\ref{comparisonsection} and Figure~\ref{lineprofiles}). We have run calculations both including and ignoring the local contributions to the background, and see negligible difference in the results. In part this is due to the fact that the background radiation temperatures are only $T_\mathrm{rad}=13-25$ K at the wavelengths of the detected FIR lines, while the typical excitation temperatures for the best-fitting models are $T_\mathrm{ex}\approx100$ K and $T_\mathrm{ex}\sim500$ K for the ME and HE transitions, respectively. In addition, most of the lines are optically thick for the best-fitting models, and hence the CO is insulated from the external radiation field.

\subsubsection{Parameter Limits}

We explore two component fits to the FIR CO emission over a large volume of 8-dimensional parameter space, applying physical limits to the model parameters. The most important prior restrictions are placed on the velocity gradient. For self-gravitating clouds in virial equilibrium, we can approximate $(dv/dr)_\mathrm{vir} \approx 10$ km\,s$^{-1}$\,pc$^{-1}$ (${n_\mathrm{H}}_2$/$10^5$ cm$^{-3}$)$^{1/2}$~\citep{Goldsmith2001}. The actual velocity gradient may be larger due to additional sources of gravitational potential, a high pressure inter-cloud medium, or non-virialized motion~\citep{Bryant1996}, but smaller values are unlikely. Defining $K_\mathrm{vir}$ as the ratio between $dv/dr$ and $(dv/dr)_\mathrm{vir}$~\citep{Papadopoulos2007}:

\begin{equation}
K_\mathrm{vir} = \frac{dv/dr}{\mathrm{10\,\,km\,s^{-1}\,pc^{-1}}} \biggl( \frac{{n_\mathrm{H}}_2}{\mathrm{10^5\,\,cm^{-3}}} \biggr)^{-1/2},
\label{kvirdef}
\end{equation}

\noindent we restrict parameter space to $K_\mathrm{vir} \ge 1$. The largest measured velocity gradient in NGC 1068 is in the H$_2$O maser disk associated with the AGN. The line of sight velocities of the maser spots shift by $\sim$$600$ km\,s$^{-1}$ over a $\sim$$2$ pc linear range~\citep{Gallimore2001}, corresponding to an effective $dv/dr\sim300$ km\,s$^{-1}$\,pc$^{-1}$. To accommodate the maser disk and other high dispersion regions in our models, we extend our calculations up to $dv/dr = 1000$ km\,s$^{-1}$\,pc$^{-1}$. We note that restricting $K_\mathrm{vir} \ge 1$ and $dv/dr \le 1000$ km\,s$^{-1}$\,pc$^{-1}$ combine to limit the density to ${n_\mathrm{H}}_2 \le 10^9$ cm$^{-3}$, but as we discuss below, such high densities are ruled out by other considerations. We calculate the total gas mass in our models as $M_\mathrm{gas} = 1.36\times {M_H}_2$, including the contribution from helium. \citet{Schinnerer2000} estimate a dynamical mass of $M_\mathrm{dyn}=9\times10^8$ $M_\odot$ for the CND, and we discard any of our models in which the total $M_\mathrm{gas}$ of the two components exceeds $M_\mathrm{dyn}$.

The range of parameter space allowed by the CO data can be further reduced by considering the H$_2$ pure rotational lines, which arise from states with similar upper energy levels as the FIR CO transitions. We calculate the H$_2$ rotational spectrum for each model, under the simplifying assumption that the lines are optically thin and thermalized, and the H$_2$ ortho/para ratio is thermalized. We then rule out any model that overpredicts the flux in any of the lines measured in the large ($14\arcsec - 27\arcsec$) ISO-SWS apertures~\citep{Lutz2000all}. These prior constraints on the LVG model parameters are summarized in Table~\ref{lvgrestrictions}.

\begin{table}
\begin{center}
\caption{LVG Model Restrictions\label{lvgrestrictions}}
\begin{tabular}{lllllllllllllllllllllllllllccccccccccccrrrrr}
\tableline\tableline
1) $K_\mathrm{vir} \ge 1$      \\
2) $dv/dr \le 1000$ km\,s$^{-1}$\,pc$^{-1}$    \\
3) $1.36\times[{M_H}_2(\mathrm{ME}) + {M_H}_2(\mathrm{HE})] \le 9\times10^8$ $M_\odot$   \\
4) H$_2$ rotational lines not overproduced                                   \\
\tableline
\tableline
\end{tabular}
\end{center}
\end{table}

\subsubsection{General Features of Good-Fitting Models} \label{goodfits}

We proceed by generating model SEDs over a regular 8-dimensional grid, and for each model calculating $\chi^2$ in the normal manner. With 11 data points and 8 free parameters, our modeling has 3 degrees of freedom (dof). Here we discuss the general properties of the set of solutions for which $\chi^2 - \chi^2_\mathrm{min} \le 1$, corresponding to $\chi^2$/dof $\le1.1$. In Figure~\ref{coSEDLVG} we show the range of SEDs covered by this set of good solutions, and for the single best fit model we show the decomposition into the ME and HE components. The CO(18-17) and CO(19-18) lines typically receive comparable contributions from the ME and HE components, while the lower- and higher-$J$ transitions are dominated by the ME and HE components, respectively. The shape of the ME SED is relatively well constrained, and peaks in the $J_\mathrm{upper} = 13-16$ range. The single dish measurements of CO(1-0), CO(2-1), CO(3-2), and CO(4-3) we show in Figure~\ref{coSEDLVG} were obtained with $11\arcsec-21\arcsec$ beams, in some cases comparable to the Herschel-PACS resolution~\citep{Israel20091068}. However, these low-$J$ lines receive strong contributions from the lower excitation gas in the $\approx$$15\arcsec$ radius starburst ring, and do not constrain our models. The interferometric measurements of the CO(1-0) flux in the CND range from $20-120$ Jy\,km\,s$^{-1}$~\citep{Schinnerer2000,Krips2011}, larger than the median ME model flux of 7 Jy\,km\,s$^{-1}$, suggesting the ME component contributes no more than a minor fraction of the observed low-$J$ emission. The HE component is more poorly constrained at the high-$J$ end, and is well fit by SEDs peaking from $J_\mathrm{upper}=25$ up to $J_\mathrm{upper}=35$. For these latter models, our upper limits to CO(28-27) and CO(34-33) become useful constraints.

The FIR CO emission is an important coolant of the nuclear molecular ISM, but does not dominate. The total luminosity emitted in the 11 transitions detected here is $L_\mathrm{CO,FIR} = 3.3 \times 10^6$ $L_\odot$, and summing the modeled ME and HE emission over all transitions yields $L_\mathrm{CO,ME} + L_\mathrm{CO,HE} = (5.7-10.2) \times 10^6$ $L_\odot$. For the highest excitation models some additional cooling may arise from the $J_\mathrm{upper} > 40$ transitions not included in our LVG calculation, and significant emission is also expected in the $J_\mathrm{upper} \le 13$ submillimeter transitions. The total emission in the H$_2$ 0-0 S(1), S(3), S(4), S(5), and S(7) rotational lines detected by ISO-SWS is $L_\mathrm{H_2} = 1.7\times10^7$ $L_\odot$~\citep{Lutz2000all}. Treating the upper limits to S(0), S(2), and S(9) as detections increases this by a factor of $1.5$, while at the same time some fraction of the lowest-$J$ emission measured with the largest apertures may arise from the starburst ring. Our PACS scans have also detected a number of OH and H$_2$O transitions. The bulk of the emission in these molecules detected in the central spaxel likely arises from the unresolved CND, and with this assumption we estimate nuclear luminosities of $L_\mathrm{OH,FIR} \approx 1.5 \times 10^7$ $L_\odot$ and $L_\mathrm{H_2O,FIR} \approx 3.0 \times 10^6$ $L_\odot$. The FIR range includes the strongest OH lines at 79 $\mu$m, 119 $\mu$m, and 163 $\mu$m, while for H$_2$O (as with CO) the longer wavelength emission should be strong. In the FIR the CO and H$_2$O cooling is therefore comparable, while the FIR CO luminosity is weaker by a moderate factor than the OH and H$_2$ rotational emission. 

\subsubsection{Bayesian Analysis} \label{bayesian}

We follow the Bayesian approach outlined by~\citet{Ward2003} to quantify the probable values of the model parameters. We consider an 8-dimensional array of bins centered on the grid points for which we have generated a model SED, and calculated a $\chi^2$ value. The probability that the actual parameter set falls within a given bin is proportional to the product of bin size, likelihood $L \propto \mathrm{exp(-\chi^2)}$, and an assumed prior probability. We choose priors that are flat in the logarithm of each parameter, and that go to zero for any model that violates one of the restrictions listed in Table~\ref{lvgrestrictions}.

\begin{figure}
\epsscale{1.1}
\plotone{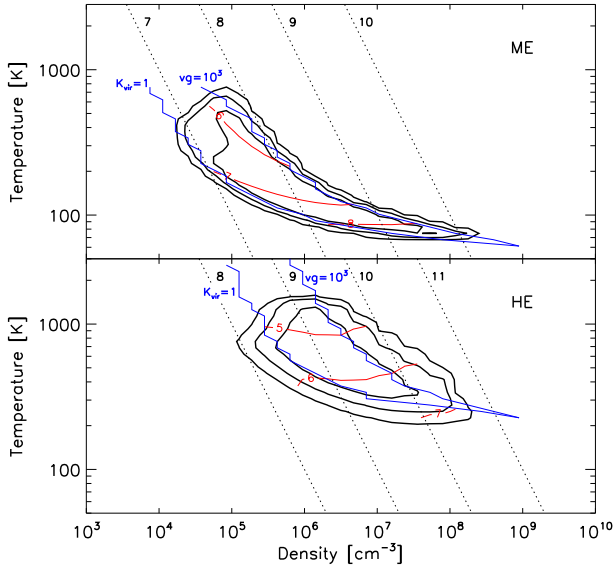}
\caption{\textit{Top:} Joint density-temperature probability density function for the ME component. Contours are drawn at 68\%, 95\%, and 99\% enclosed probability. The mean of log(${M_H}_2/M_\odot$) needed to reproduce the absolute line fluxes is shown by the red curves, and the logarithm of the thermal pressure (log(${n_\mathrm{H}}_2\times T_\mathrm{kin}$) in K cm$^{-3}$) is shown by the dotted black lines. Blue curves show the density-temperature relation giving the best solution for $K_\mathrm{vir}=1$ and $dv/dr=10^3$ km\,s$^{-1}$\,pc$^{-1}$ (see text). \textit{Bottom:} Same as the top, but for the HE component.\label{chisqrdplot}}
\end{figure}

In Figure~\ref{chisqrdplot} we show the joint density-temperature probability density functions for both components, with contours at 68\%, 95\%, and 99\% of the enclosed probability. The close similarity between the 95\% and 99\% (and in some regions also the 68\%) contours is due to a truncation of the density function following the violation of one of our model restrictions. Starting with the ME component, the behaviour of the contours may be understood as follows. For either a fixed $K_\mathrm{vir}$ or $dv/dr$, the shape of the CO SED is approximately conserved if an increase in density is matched by an approriate decrease in temperature. The pair of blue curves in Figure~\ref{chisqrdplot} show the density-temperature relation best fitting the data for $K_\mathrm{vir}=1$ and $dv/dr = 1000$ km\,s$^{-1}$\,pc$^{-1}$. As the shape of the ME SED is well constrained, the region of acceptable parameter space for either set of models corresponds to a narrow band centered on the best fit curve. These two bands intersect at ${n_\mathrm{H}}_2=10^9$ cm$^{-3}$ and diverge at lower densities, thereby bracketing a region of high probability filled by models with intermediate velocity gradients. For lower temperatures a decreasing fraction of the CO is excited to the higher-$J$ states, and a larger total mass is needed to reproduce the absolute line fluxes. For ${n_\mathrm{H}}_2\gsim10^8$ cm$^{-3}$ the gas mass exceeds the dynamical mass, and consequently a small region of parameter space is excluded. For higher temperatures the model emission does not fall off with increasing $J$ in the $J_\mathrm{upper} \approx 18$ region as rapidly as the data require, and the lower quality fits limit the extent of the 68\% confidence interval above $T_\mathrm{kin} \approx 230$ K. For $T_\mathrm{kin} \gsim 400$ and ${n_H}_2 \lesssim 10^5$ cm$^{-3}$ the models begin to overproduce the H$_2$ rotational line detections and upper limits, and this accounts for the sharp cutoff in the probability density in the upper left region.

The shape of the HE SED is not as well constrained as the ME SED, and the H$_2$ rotational lines provide a stronger constraint to the extent of the probability density contours in Figure~\ref{chisqrdplot}. The detection of CO(30-29) excludes SEDs peaking at $J_\mathrm{upper}\lesssim25$, but the upper limits at $J_\mathrm{upper}\ge 34$ provide a softer restriction on higher pressure models peaking at $J_\mathrm{upper} \gsim 30$. As a result, the $dv/dr \le 1000$ km\,s$^{-1}$\,pc$^{-1}$ restriction no longer forms the upper right boundary of the density function, and models are allowed to extend into this higher density and temperature region until the modeled H$_2$ emission exceeds the observations. The measured S(1) upper limit provides the most important constraint at high densities (${n_H}_2 \approx 10^{8}$ cm$^{-3}$) and low temperatures ($T_\mathrm{kin} \approx 300$ K). The higher excitation H$_2$ lines become more important at higher temperatures, and the S(7) transition limits the upper left region at $T_\mathrm{kin}\approx1000$ K.

\begin{figure}
\epsscale{1.1}
\plotone{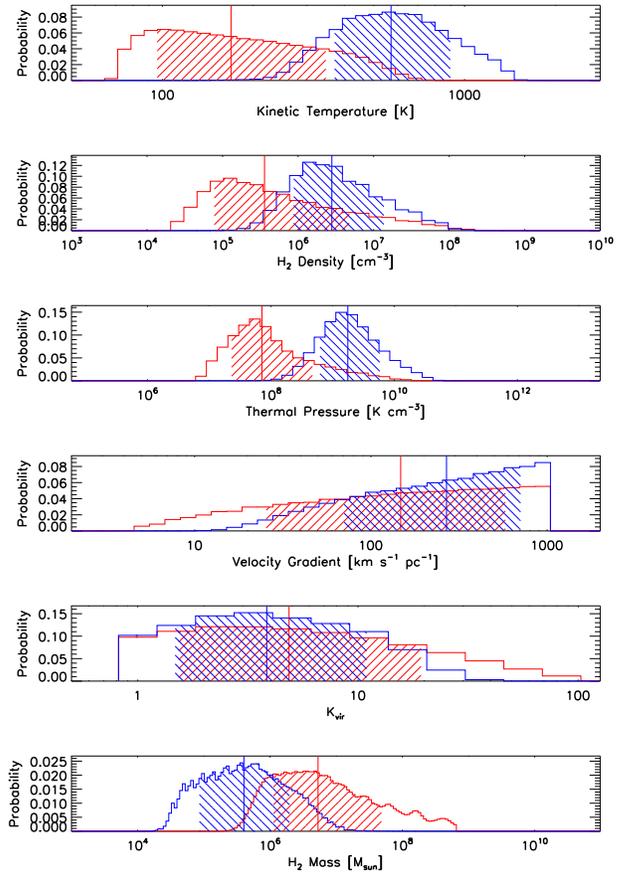}
\caption{Probability density functions for the 4 primary model parameters, as well as for $K_\mathrm{vir}$ and the thermal pressure $P/k_B = {n_H}_2\times T_\mathrm{kin}$. In each panel the ME distribution is shown in red, and the HE in blue. The binwidths are proportional to the logarithm of the model parameter, and the ordinate indicates the probability that the actual value lies in the given bin. The vertical lines indicate the median of each distribution, and the hatched regions indicate the symmetric 68\% confidence interval.\label{qpdf}}
\end{figure}

In Figure~\ref{qpdf} we show the fully marginalized distribution functions for each of the 4 primary model parameters, as well as for $K_\mathrm{vir}$ and the thermal pressure $P/k_B = {n_H}_2\times T_\mathrm{kin}$. The distribution functions for the HE $T_\mathrm{kin}$, ${n_H}_2$, and $P/k_B$ are shifted to higher values than for the corresponding ME parameters, although the density distributions for the two components contain significant overlap. The lower limit imposed on $K_\mathrm{vir}$ translates into a lower limit to $dv/dr$ that increases with density (see equation~\ref{kvirdef}). As such, lower density solutions are found over a broad range of velocity gradients, while higher density models are limited to larger values of $dv/dr$. This accounts for the positive slope of the $dv/dr$ distribution. Similarly, the upper limit placed on $dv/dr$ generates a negative slope in the $K_\mathrm{vir}$ distribution. As the distribution functions for these two parameters are heavily influenced by our prior restrictions, our modeling produces no meaningful constraint on the dynamical state of the gas.

We take the median of each distribution as the single best estimate of the parameter value, and indicate this number with a vertical line in each panel of Figure~\ref{qpdf}. Most of the distribution functions are relatively symmetric, and we calculate the equivalent $1$$\sigma$ uncertainties in the parameter estimation by finding the range of values symmetrically enclosing 68\% of the total probability. This range is shown as the hatched area under each distribution. The results of this analysis are summarized in Table~\ref{lvgtable}. The asymmetry in the ME $T_\mathrm{kin}$ distribution results from the limit to low temperature, high density parameter space following the $K_\mathrm{vir} \ge 1$ and $M_\mathrm{gas} \le M_\mathrm{dyn}$ restrictions (the lower right region in the top panel of Figure~\ref{chisqrdplot}). For this parameter, as well as for $dv/dr$ and $K_\mathrm{vir}$, we extend the acceptable range listed in Table~\ref{lvgtable} to the truncated edge of the distribution.

\renewcommand{\thefootnote}{\alph{footnote}}
\begin{table*}
\begin{center}
\caption{LVG Model Results\label{lvgtable}}
\begin{tabular}{ccccccccccccrrrrr}
\tableline\tableline
                                  & \multicolumn{2}{c}{ME}                   & \multicolumn{2}{c}{HE}                   \\
Parameter                         & Median      & 68\% Range                 & Median      & 68\% Range                 \\
\tableline
$T_\mathrm{kin}$ [K]              & $169$       & $71\tablenotemark{a}-347$  & $571$       & $372-896$                  \\
${n_H}_2$ [cm$^{-3}$]             & $10^{5.6}$  & $10^{4.9}-10^{6.7}$        & $10^{6.4}$  & $10^{5.9}-10^{7.1}$        \\
$P/k_B$ [K\,cm$^{-3}$]            & $10^{7.9}$  & $10^{7.4}-10^{8.7}$        & $10^{9.2}$  & $10^{8.8}-10^{9.8}$        \\
$dv/dr$ [km\,s$^{-1}$\,pc$^{-1}$] & $148$       & $25-1000\tablenotemark{a}$ & $269$       & $71-1000\tablenotemark{a}$ \\
$K_\mathrm{vir}$                  & $4.9$       & $1\tablenotemark{a}-19$    & $3.9$       & $1\tablenotemark{a}-11$    \\
${M_H}_2$ [$M_\odot$]             & $10^{6.7}$  & $10^{6.1}-10^{7.7}$        & $10^{5.6}$  & $10^{4.9}-10^{6.3}$        \\
\tableline
\end{tabular}
\tablenotetext{1}{Extended beyond the formal 68\% confidence interval to the truncated edge of the distribution.}
\end{center}
\end{table*}
\renewcommand{\thefootnote}{\arabic{footnote}}

\subsubsection{Varying the CO Abundance} \label{reducedxco}

In the above analysis we assumed a CO abundance of [CO/H$_2$] $ = 10^{-4}$, motivated by abundance measurements in Galactic molecular clouds~\citep[e.g., $\mathrm{[}$CO/H$_2$$\mathrm{]}$ $ = 8.5\times10^{-5}$;][]{Frerking1982}. However, the CO abundance in the center of NGC 1068 may be different. Noting the general trend of higher metallicities in galactic centers, \citet{Israel20091068} adopts an elevated carbon abundance for the center of NGC 1068, and derives [CO/H$_2$] $ = 4\times10^{-4}$. At the same time, in molecular clouds exposed to intense X-ray fields (as we expect for NGC 1068; see sections~\ref{meheating} and~\ref{heheating}), the CO abundance may be significantly reduced~\citep{Krolik1989,Maloney1996}. In the models of~\citet{Meijerink2005} with high ratios of incident X-ray flux to gas density, the bulk of the gas phase carbon is not bound up in CO until large depths ($N_H\gsim10^{23.5}$ cm$^{-2}$) into the cloud. This yields a large column of warm gas with reduced CO abundance that may contribute to the FIR CO line emission.

\begin{figure}
\epsscale{1.1}
\plotone{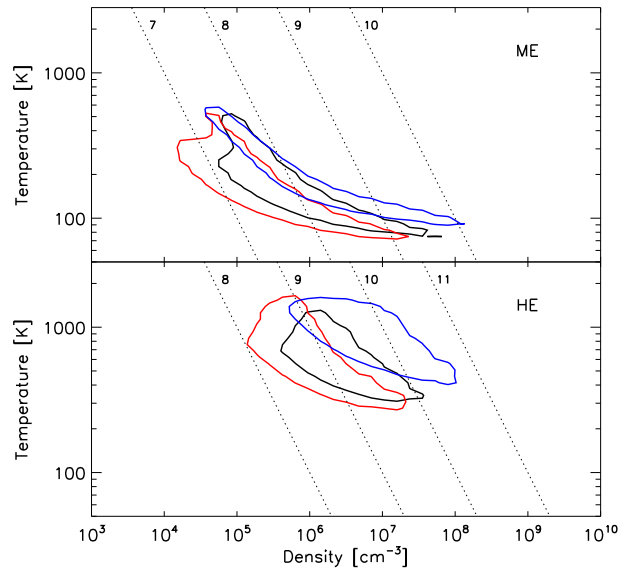}
\caption{
\textit{Top:} Joint density-temperature probability density function for the ME component, as in Figure~\ref{chisqrdplot}, but for various CO abundances. Contours are drawn at 68\% enclosed probability for [CO/H$_2$] $ = 10^{-5}$ (blue), $10^{-4}$ (black), and $4\times10^{-4}$ (red). \textit{Bottom:} Same as the top, but for the HE component.
\label{varyingxcoplot}}
\end{figure}

To explore the effects of an altered CO abundance, we have repeated our LVG analysis for a range of [CO/H$_2$] values. In Figure~\ref{varyingxcoplot} we show the joint density-temperature probability density functions for models with [CO/H$_2$] $ = 10^{-5}$, $10^{-4}$, and $4\times10^{-4}$. For a fixed value of $dv/dr$, the line opacities scale as the product of ${n_H}_2$ and [CO/H$_2$]. The shape of the CO SED is therefore approximately conserved if an increase in [CO/H$_2$] is matched by a decrease in ${n_H}_2$, and this accounts for the shift to lower densities for a larger CO abundance. Increasing the CO abundance also leads to a reduction in the H$_2$ mass needed to maintain the absolute CO fluxes, and hence a smaller set of predicted H$_2$ rotational line fluxes. The restriction that the modeled H$_2$ emission not exceed the measured emission then provides a weaker constraint, and leads to an increase in the allowed volume of density-temperature parameter space, particularly in the high temperature region. For [CO/H$_2$] $=4\times10^{-4}$, the combination of these two effects results in only a small reduction in the derived density and H$_2$ mass, and a small increase in the derived temperature (Figure~\ref{varyingxcoplot}), and does not significantly affect the physical parameter estimates obtained previously.

If the CO abundance is reduced by a strong X-ray flux, models indicate that the hydrogen will also be largely atomic~\citep{Maloney1996,Meijerink2005}. We modify our LVG calculation to account for an atomic/molecular mixture by changing [CO/H$_2$] $\rightarrow 2$[CO/H], ${n_H}_2\rightarrow n(\mathrm{H})/2$, and ${M_H}_2\rightarrow M(\mathrm{H})$, and introducing the molecular fraction as $f_\mathrm{mol} = $ [H$_2$/H]. In each of these expressions, H is taken to represent the total number or mass of hydrogen nuclei in both atomic and molecular form. For collisional excitation of CO by atomic hydrogen, we adopt the same rate coefficients as for excitation by H$_2$~\citep[see the discussion in][]{Flower2010}. For simplicity, we also assume that any reduction in the nominal CO abundance is matched by an identical reduction in the molecular fraction, and set $f_\mathrm{mol} = 2$[CO/H]$ \times 10^4$. Following the same reasoning as discussed above in the context of an increased CO abundance, a reduction in the CO abundance will increase the derived density and total mass. However, as the CO to H$_2$ ratio is conserved, the predicted H$_2$ line fluxes are to first order unchanged. The restriction that the modeled H$_2$ emission not exceed the measured emission then provides a comparable constraint on the allowed volume of density-temperature parameter space as with the nominal CO abundance. For the ME component, using 2[CO/H] $ = 10^{-5}$ shifts the solutions to only moderately higher densities, temperatures, pressures, and masses (Figure~\ref{varyingxcoplot}). For the HE component, the increase in these parameters is more significant, although the mean values of each parameter remain within the 68\% range for our nominal [CO/H$_2$] $ = 10^{-4}$ model (Table~\ref{lvgtable}).

\section{Comparison with Other Molecular Tracers} \label{comparisonsection}

\subsection{Physical Parameters and Mass Fraction} \label{physicalparameters}

The bulk of the emission traced by high resolution millimeter and near-IR molecular gas maps in the central 10$\arcsec$ arises from the $\sim$$5\arcsec$ ($\sim$$350$ pc) CND~\citep{Schinnerer2000,Galliano2002,GarciaBurillo2010}. Our LVG modeling indicates that no more than half of the CO(1-0) emission in the CND is generated by our ME and HE components, indicating lower excitation material is present. The physical conditions of this low excitation component have been studied by many groups. \citet{Tacconi1994} detected strong CO(4-3) emission toward the center of NGC 1068, and combined this with an interferometric CO(1-0) measurement to show that the gas was both warm ($T_\mathrm{kin} \ge 70$ K) and dense (${n_H}_2\ge2\times10^4$ cm$^{-3}$). Subsequent modeling of $J_\mathrm{upper} \le 4$ transitions of $^{12}$CO and $^{13}$CO have typically adopted a fixed $T_\mathrm{kin}=50$ K, and derived ${n_H}_2\sim10^4 - 10^5$ cm$^{-3}$~\citep{Sternberg1994,Helfer1995,Usero2004}. \citet{Krips2011} have obtained fluxes of the lowest 3 transitions of both $^{12}$CO and $^{13}$CO with $\lesssim2\arcsec$ resolution. For a subsection of the CND in which the gas appears perturbed by the radio jet, and hence possibly shock-heated,~\citet{Krips2011} derive $T_\mathrm{kin} \ge 200$ K and ${n_H}_2=10^{3.5-4.5}$ cm$^{-3}$. While this section may not be representative of the CND as a whole, this analysis does suggest that the global temperatures may be higher than assumed by previous authors, and similar to that derived for our ME component. We also note that~\citet{Kamenetzky2011} have similarly derived a globally high temperature ($T > 100$ K) for the CND based on an analysis of CS and other high density tracers. Given this range of temperature estimates, the clearest difference between the ME CO component and the lower excitation material traced by the low-$J$ CO lines is the higher density ($n_{H_2} \sim 10^{5.6}$ cm$^{-3}$ vs $n_{H_2} \lesssim 10^5$ cm$^{-3}$). This suggests a scenario in which the FIR lines trace denser material in the CND, which coexists with a more diffuse medium that generates the millimeter CO emission.

The mass fraction of the high excitation gas may be estimated by comparing the ME and HE masses derived here with the mass traced by CO(1-0). With a standard Galactic conversion factor $N(H_2)/I_\mathrm{CO} = 2\times10^{20}$ cm$^{-2}$\,(K\,km\,s$^{-1}$)$^{-1}$,~\citet{Schinnerer2000} estimate a mass of ${M_H}_2 = 5\times10^7$ $M_\odot$ for the CND. This is similar to previous estimates from~\citet[][after correcting their mass to the $d=14.4$ Mpc used here]{Planesas1991} and~\citet{Helfer1995}, both of which used similar conversion factors. However,~\citet{Usero2004} have derived a much lower $N(H_2)/I_\mathrm{CO} = 0.3\times10^{20}$ cm$^{-2}$\,(K\,km\,s$^{-1}$)$^{-1}$ (for the [CO/H$_2$] $ = 10^{-4}$ used here) from their excitation modeling of the low-$J$ CO emission from the CND. \citet{Papadopoulos1999CO} and~\citet{Israel20091068} have also derived conversion factors significantly lower than the Galactic value by analyzing the low-$J$ CO emission averaged over beam sizes of $\sim$$1\arcmin$ and $\sim$$21\arcsec$, respectively. These authors have suggested the lower conversion factor arises from the gas not being virialized. At the same time,~\citet{Krips2011} have reported a CO(1-0) flux from the central $4\arcsec$ of 120 Jy\,km\,s$^{-1}$, much larger than the 20 Jy\,km\,s$^{-1}$ value reported by~\citet{Schinnerer2000}. For $N(H_2)/I_\mathrm{CO} = 0.3\times10^{20}$ cm$^{-2}$\,(K\,km\,s$^{-1}$)$^{-1}$, and $F_\mathrm{CO(1-0)} = 20 - 120$ Jy\,km\,s$^{-1}$, we estimate ${M_H}_2 = (0.5-3) \times10^7$ $M_\odot$. This is comparable to the ${M_H}_2 = (0.1-5) \times10^7$ $M_\odot$ range we associate with our ME component. While the uncertainties of both numbers are high, this suggests that the ME component makes a non-negligible contribution to the total mass budget. The HE emission traces a lower mass of warmer gas, albeit still cooler than the small amount ($M\sim10^3$ $M_\odot$) of hot ($T\sim2000$ K) gas detected in the near-IR H$_2$ lines~\citep{Rotaciuc1991,Blietz1994,Galliano2002}.

\subsection{Line Profiles} \label{linekinematics}

\subsubsection{PACS Lines} \label{pacslinekinematics}

\begin{figure}
\epsscale{1.1}
\plotone{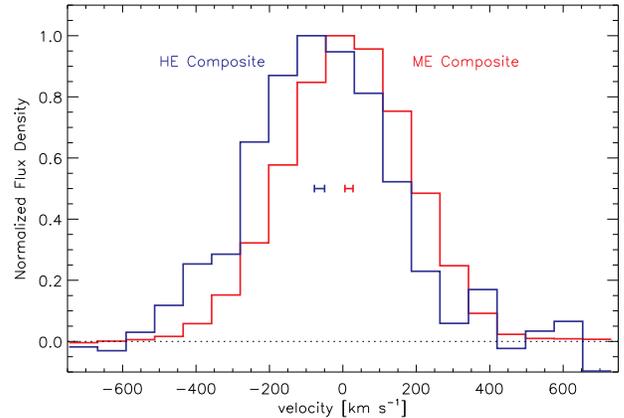}
\caption{Average profiles of the ME (red) and HE (blue) CO lines. Prior to averaging, each line is smoothed to a resolution of 311 km\,s$^{-1}$. The horizontal error bars shown underneath the centroid of each profile indicate the estimated $\pm1\sigma$ calibration uncertainty of the stacked spectra.\label{MEHEcomposite}}
\end{figure}

Further insight into the nature of the ME and HE components may be obtained by comparing the FIR CO line profiles with those of other molecular tracers. In particular, we seek to understand the physical origins of the velocity shift between the two components. To better demonstrate the spectral shift, in Figure~\ref{MEHEcomposite} we show the composite spectra obtained by averaging the ME ($J_\mathrm{upper}=14-17$) and HE ($J_\mathrm{upper}=20-22,24,30$) profiles. The PACS spectral resolution changes from $191-246$ km\,s$^{-1}$ and $127-311$ km\,s$^{-1}$ over the wavelength range corresponding to the ME and HE transitions, respectively. The composite spectra have been obtained by smoothing each line to 311 km\,s$^{-1}$ resolution (each line measured with instrumental resolution $\delta v$ is smoothed with a Gaussian kernel of $\mathrm{FWHM} = \sqrt{311^2 + \delta v^2}$) and resampling to a common velocity grid. Gaussian fits to the composite profiles yields similar central velocities ($V_\mathrm{ME} = 17$ km\,s$^{-1}$ and $V_\mathrm{HE} = -64$ km\,s$^{-1}$) as obtained from averaging the central velocities of the individual lines ($V_\mathrm{ME} = 17$ km\,s$^{-1}$ and $V_\mathrm{HE} = -59$ km\,s$^{-1}$; Figure~\ref{coSEDLVG}). The uncertainties in the centroids of the composite spectra are dominated by the spectral calibration uncertainties in the individual lines. With each line having a calibration error of $\sigma_i$, we estimate the error ($\sigma$) in the composite spectra as $\sigma^2 = <\sigma_i^2>/N$. This gives $\sigma_\mathrm{ME}=11$ km\,s$^{-1}$ and $\sigma_\mathrm{HE}=14$ km\,s$^{-1}$, and we show these as horizontal error bars in Figure~\ref{MEHEcomposite}.

\begin{figure}
\epsscale{1.1}
\plotone{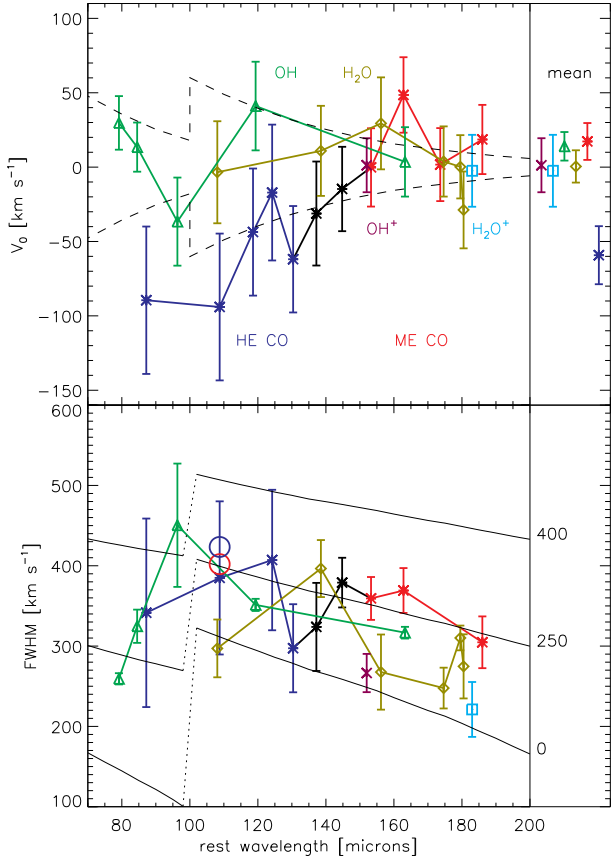}
\caption{\textit{Top:} Measured line centroids for the CO, OH (green), H$_2$O (brown), OH$^+$ (purple), and H$_2$O$^+$ (cyan) lines detected in our PACS range scans. The CO lines are color coded by ME (blue), HE (red), and intermediate (black). Each OH point is an average of a doublet, and the OH$^+$ and H$_2$O$^+$ points represent averages over all detected fine-structure lines. The dashed line shows the instrumental wavelength shift induced by pointing offsets of $\pm$$2\arcsec$ in the dispersion direction. On the right hand side we show the mean centroid of each tracer. \textit{Bottom:} Same as the top, but showing the measured FWHM of each line. At $\lambda=109$ $\mu$m we show the linewidths of the ME and HE composite profiles as open symbols. Solid curves show the expected measured widths for lines with intrinsic widths of 0, 250, and 400 km\,s$^{-1}$.\label{velcentroid}}
\end{figure}

In addition to CO, the PACS scans detect strong molecular emission from several transitions of OH and H$_2$O, and weaker emission from OH$^+$, H$_2$O$^+$, and other molecules. Some of the OH and H$_2$O lines show extended emission over the PACS array, but for each line the flux in the central spaxel is dominated by a compact source that we associate with the CND. The OH$^+$ and H$_2$O$^+$ emission is also unresolved, although an association of these lines with the same molecular gas is less certain. In Figure~\ref{velcentroid} we compare the central velocities and widths of these lines. For OH, each point represents the average properties of a doublet, while for OH$^+$ and H$_2$O$^+$ each point is an average of multiple fine-structure transitions. On the right hand side of the top panel we show the mean central velocity of each molecule, with the ME and HE CO components separately. The HE CO is the only tracer systematically offset from the systemic velocity. In the bottom panel we show the widths of the ME and HE composite profiles as open points at $\lambda = 109$ $\mu$m, where the smoothed 311 km\,s$^{-1}$ resolution is equal to the instrumental resolution. Overplotted are a set of curves showing the expected measured linewidths for intrinsic linewidths of 0, 250, and 400 km\,s$^{-1}$. The individual CO lines, as well as the composite profiles, are consistent with an intrinsic $\mathrm{FWHM}\sim250$ km\,s$^{-1}$ for both the ME and HE components. The other molecular lines are somewhat narrower.

The $50-80$ km\,s$^{-1}$ offset between the HE lines and the other molecular tracers in Figure~\ref{velcentroid} is $\lesssim$$1/3$ of the PACS spectral resolution, and at this level instrumental effects must be considered. The PACS spectral response depends on the illumination of the slit, such that a physical translation of a source on the sky in the dispersion direction will produce a wavelength shift in the spectral profile~\citep{Poglitsch2010}. In the top panel of Figure~\ref{velcentroid} we show the equivalent velocity shift resulting from moving a point source $\pm2\arcsec$ from the center of the slit. A comparison with the CO velocities shows that a $\sim$$2-3\arcsec$ offset between the centers of the slit and the HE CO emission could be partially responsible for the measured wavelength shifts. With the exception of CO(17-16), CO(24-23), and CO(30-29), each point in Figure~\ref{velcentroid} corresponds to a line measured in the same set of concatenated range scans. Therefore if all of the molecular emission is presumed to arise from the same region on the sky, a wavelength shift in the HE CO lines induced by a pointing offset should be matched by a comparable shift in the $\lambda\approx100-150$ $\mu$m OH and H$_2$O lines, and this is not observed. Alternatively, the HE CO lines may be modeled as arising from a region centered $\sim$$2-3\arcsec$ away from the source of the OH and H$_2$O emission. However, the CO(30-29) line at $\lambda=87$ $\mu$m was observed separately from the series of concatenated scans, with a slit position angle differing by $\approx$$180\arcdeg$. Any pointing-induced shift to the CO(30-29) line should then have a comparable magnitude but be in the opposite direction of the shifts in the other HE CO lines, and this is also not the case. We therefore argue that the observed wavelength shift represents a real velocity offset, with instrumental effects playing no more than a minor role.

\subsubsection{High Resolution Observations} \label{highreskinematics}

To search for kinematic substructure in the CND that might explain the velocity shifts among the FIR CO lines, we compare the FIR line profiles with those of high resolution observations of CO(2-1)~\citep{Schinnerer2000} and H$_2$ 1-0 S(1)~\citep{MuellerSanchez2009}. In Figure~\ref{lineprofiles}a we show maps of these two tracers in the central $4\arcsec\times6\arcsec$. At $0.7\arcsec$ resolution the CO(2-1) emission from the CND is dominated by a pair of knots centered $\sim$$1\arcsec$ east and $\sim$$1.5\arcsec$ west of the AGN, connected by lower surface brightness emission. Similar morphology is also seen in the $\sim$$1\arcsec$ resolution aperture synthesis maps of CN and SiO~\citep{GarciaBurillo2010} and of HCN and HCO$^+$~\citep{Krips2011}. Observations of the H$_2$ 1-0 S(1) line at 0.075$\arcsec$ resolution have resolved the CND into a $\sim$$1.2\arcsec$ radius ring (hereafter the "H$_2$ ring") centered $\sim$$0.6\arcsec$ southwest of the AGN. The eastern knot is also prominent in the H$_2$ map, while the western knot is much fainter. The H$_2$ map also shows strong emission from a clump $\sim$$1\arcsec$ to the north of the AGN that is not prominent in the CO(2-1) map.

\begin{figure*}
\epsscale{1.1}
\plotone{figure08.epsi}
\caption{\textbf{a)} CO(2-1) (contours) and H$_2$ 1-0 S(1) (color) images of the central $4\arcsec\times6\arcsec$ from~\citet{Schinnerer2000,MuellerSanchez2009}, \textbf{b)} blue and red components of CO(2-1), with crosses marking the centers of the apertures in a), \textbf{c)} H$_2$ 1-0 S(1) image of the central $0.4\arcsec$, \textbf{e-i)} smoothed H$_2$ 1-0 S(1) (black solid) and CO(2-1) (black dashed) profiles from selected apertures in the H$_2$ ring compared with the ME (red) and HE (blue) composite profiles, and $\pm40$ km\,s$^{-1}$ horizontal error bar indicating the calibration uncertainty between the FIR CO and H$_2$ spectra, \textbf{d,j)} smoothed H$_2$ 1-0 S(1) profiles from the northern and southern streamer compared with the ME and HE composite profiles, and $\pm40$ km\,s$^{-1}$ horizontal error bar.\label{lineprofiles}}
\end{figure*}

In panel b we show separate maps of the CO(2-1) emission integrated over blue and red velocities (see also Figure 6 in~\citet{Krips2011}). The strong emission to the east is largely to the blue of the systemic velocity, while the emission to the west separates into a blueshifted SW component and a redshifted NW component. The H$_2$ velocity field is generally similar, with the exception of the northern clump as discussed below. \citet{Schinnerer2000} model the CO(2-1) kinematics as a warped disk, while subsequent study of the H$_2$ and millimeter tracers has shown evidence for additional non-circular motion, possibly indicating a radially expanding component~\citep{Galliano2002,Davies2008conference,GarciaBurillo2010,Krips2011}. In panels e through i we compare the composite ME and HE spectral profiles with those of CO(2-1) and H$_2$ extracted from selected apertures. For each panel we have smoothed the CO(2-1) and H$_2$ spectra to the same resolution (311 km\,s$^{-1}$) and resampled to the same grid as used for the ME and HE composite spectra.

The most important result of this comparison is the mismatch between the FIR line profiles and the broad and highly blueshifted H$_2$ emission from the bright knot $\sim$$1\arcsec$ to the north of the AGN, shown in panel f. In the high spectral resolution H$_2$ map presented in~\citet{Galliano2002} this is the one region in the H$_2$ ring that displays a double-peaked profile (the two peaks become blended following the smoothing done here), with an extra emission component to the blue that is a clear outlier to their simple rotation plus expansion model. This region is also the site of the strongest Br$\gamma$ emission in the H$_2$ ring, which displays a broad (FWHM $\sim1000$ km\,s$^{-1}$) line. The radio jet exits the nucleus to the north~\citep{Gallimore1996paperI}, and may interact with material in the narrow-line region~\citep{Axon1998}. \citet{Galliano2002} and~\citet{Galliano2003} interpret the broad and complex Br$\gamma$ and H$_2$ line profiles as arising from gas perturbed by the jet, and additionally suggest that the ionization resulting from the jet-ISM interaction may be responsible for the strength of the Br$\gamma$ emission in this region. The smoothed H$_2$ profile in panel f is inconsistent with either the ME or HE composite spectrum, and we conclude that this region of the CND does not dominate the PACS CO emission. In sections~\ref{meheating} and~\ref{heheating} we discuss possible excitation mechanisms for the FIR CO, including shock heating. The fact that we can rule out an origin in the region of the H$_2$ ring with the best evidence for molecular gas perturbed by the jet leads us to suggest that such jet-driven shocks in the H$_2$ ring do not excite the high-$J$ CO.

Aside from the bright H$_2$ knot in the north, the ME profile is generally consistent with much of the rest of the H$_2$ ring. The best fit is with the profile of the strong H$_2$ emission from the east. The profiles from the NW and SW are moderately red and blueshifted with respect to the ME composite, but a combination of these regions, and indeed of the H$_2$ emission integrated over the entire CND (excluding the northern knot), would also generate a reasonable fit. We note that the H$_2$ linewidths (FWHM $=390-440$ km\,s$^{-1}$) are larger than the CO(2-1) linewidths (FWHM $= 330-350$ km\,s$^{-1}$) in each panel, and better match the widths of the composite ME and HE profiles (FWHM $= 400-420$ km\,s$^{-1}$). This may indicate that the hot gas probed by the H$_2$ is a better tracer of the material producing the high-$J$ CO, although due to extinction of the H$_2$ line~\citep[see, e.g., ][]{Galliano2003}, the morphology of the FIR CO emission may be different than the H$_2$ image.

The blueshift of the HE emission is more challenging to match. The H$_2$ spectra from the blue regions in the E and SW are $\approx$$65$ and $\approx$$45$ km\,s$^{-1}$ to the red, respectively, while the CO(2-1) profiles from the same regions are too narrow. However, the uncertainty in the mean centroid of the HE lines is $\approx$$20$ km\,s$^{-1}$, while the H$_2$ calibration error is smaller. Registration errors between the velocity frames of the PACS spectra and the ground-based H$_2$ spectra are also likely to be important at the level of $\sim$$10-20$ km\,s$^{-1}$, although more difficult to quantify. With a conservative estimate of $40$ km\,s$^{-1}$ uncertainty in the relative calibration of the PACS CO and the H$_2$ spectra, the HE CO emission only differs by $1.1-1.6\sigma$ with the H$_2$ emission from the E or SW regions. We conclude that while the HE emission profile is not naturally matched to the H$_2$ or CO(2-1) emission from the H$_2$ ring, an association with the bluest material to the E or SW may be within the measurement errors, and should not be excluded. 

In panel c we show the $0.025\arcsec$ resolution H$_2$ 1-0 S(1) map, which identifies two gas clouds streaming toward the nucleus on highly elliptical orbits from the north and south~\citep[hereafter the "northern" and "southern" streamers;][]{MuellerSanchez2009}. The northern streamer is connected to the H$_2$ ring in both H$_2$ emission and mid-IR continuum, and has been proposed as a means by which material is transported to the AGN~\citep{Tomono2006,MuellerSanchez2009}. The southern streamer is detected to within $\sim$$10$ pc of the AGN, and may play a role in obscuring the nuclear emission. The southern streamer is modeled to lie in front of the AGN in the plane of the galaxy, and has a redshifted velocity that increases from $\sim$$50$ km\,s$^{-1}$ to $\sim$$80$ km\,s$^{-1}$ as the gas moves to within a projected distance of $\lesssim0.1\arcsec$. The northern streamer approaches the AGN from behind, and the brightest emission occurs $0.4\arcsec$ from the center at a blueshifted velocity of $\sim-30$ km\,s$^{-1}$. In panels d and j we compare the smoothed profiles of these two regions with the PACS lines. The emission from the southern streamer is detected with lower S/N, but displays a profile reasonably consistent with that of the ME CO composite. The line profile of the northern streamer produces an excellent match to the HE composite. The centroids of the southern streamer and the HE CO composite differ by $\approx$$105$ km\,s$^{-1}$, which we argue is too large to be plausibly explained by calibration uncertainties between the two datasets. We conclude that in addition to an origin in the H$_2$ ring as discussed above, an origin of the ME and HE components with the southern and northern streamers, respectively, would be consistent with the line profiles. 

\section{Heating the ME Component} \label{meheating}

The ME CO emission arises from a warm ($T_\mathrm{kin}\sim169$ K) and dense ($n_\mathrm{H_2}\sim10^{5.6}$ cm$^{-3}$) component, and with a total mass of $M_{H_2}\sim10^{6.7}$ $M_\odot$, represents an important fraction of the ISM in the CND. The kinematic analysis presented in section~\ref{linekinematics} shows that this component may be readily attributed to the H$_2$ ring, or possibly to the southern streamer. Here we consider potential heating mechanisms, and conclude that X-ray and shock heating are both plausible, while heating by far-UV photons is less likely. We further argue that no plausible heating mechanism is consistent with an origin in the southern streamer, and hence the emission is likely associated with the H$_2$ ring.

\subsection{X-Ray Heating} \label{mexrayheating}

The AGN in NGC 1068 emits a hard X-ray luminosity of $L_\mathrm{2-10\,keV} = 10^{43-44}$ erg\,s$^{-1}$~\citep{Iwasawa1997,Colbert2002}. Our view of the AGN is obscured by a Compton-thick medium, but the extended emission detected by Chandra in the $6-8$ keV band demonstrates that the nuclear X-rays irradiate the ISM over the central $\sim$kpc~\citep{Ogle2003,GarciaBurillo2010}. Hard X-rays penetrate deeply into clouds, and efficiently heat large columns of molecular gas through photoionization heating~\citep{Maloney1996}. The bright H$_2$ 1-0 S(1) emission in the CND has been attributed to X-ray heated gas~\citep{Rotaciuc1991,Maloney1997,Galliano2002}, and X-ray heating should also produce strong emission in the FIR CO lines~\citep{Krolik1989}. 

\begin{figure}
\epsscale{1.1}
\plotone{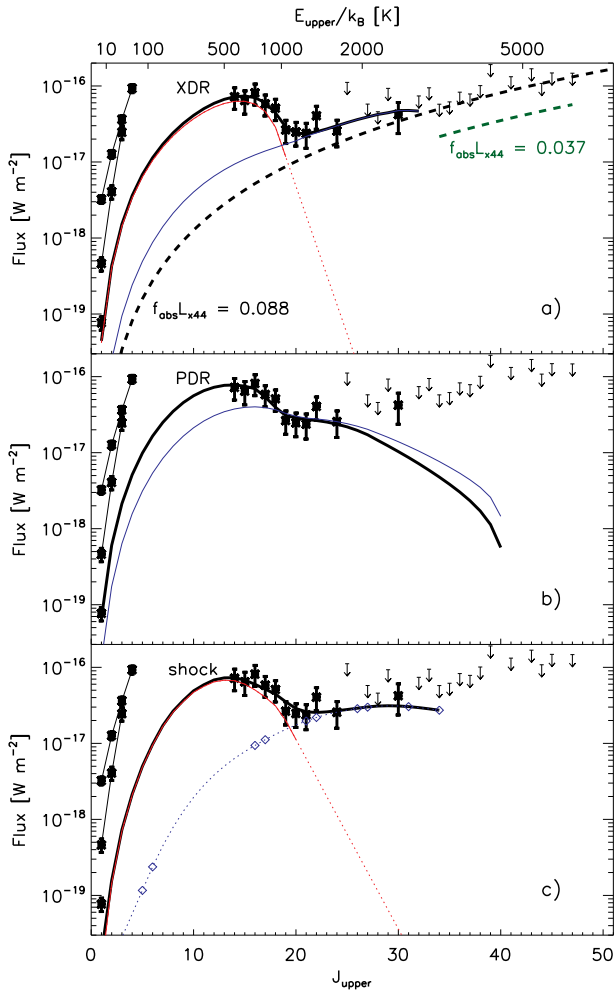}
\caption{FIR CO line fluxes and upper limits measured here, along with lower-$J$ lines from the literature (see Figure~\ref{coSEDLVG} for details), and XDR, PDR, and shock model fits. \textbf{a)} Overplotted is a two component XDR fit (solid black), with individual components in red and blue. Dotted section of the red curve shows an extrapolation of the model. Upper limits to the~\citet{Krolik1989} torus model are shown by the black (constrained by the CO[44-43] limit) and green (constrained by the stacked limit to selected $J_\mathrm{upper}=34-47$ transitions) dashed curves. \textbf{b)} Black curve shows a PDR fit to the full CO SED, and blue curve shows a separate fit to the $J_\mathrm{upper} \ge 19$ lines. \textbf{c)} Two component shock fit with same color scheme as panel a). Blue curve shows a model interpolated and extrapolated from a finite number of transitions (diamonds). Model parameters and references are discussed in the text and summarized in Table~\ref{heatingtable}. \label{coSED}}
\end{figure}

\citet{Meijerink2005} and~\citet{Meijerink2007} present a detailed photochemical modeling of X-ray dominated regions (XDRs) that includes predictions for the emergent CO line intensities as a function of the gas density ($n_\mathrm{H}$) and incident hard X-ray flux ($F_X = F_\mathrm{2-10\,keV}$). We use their type A models, which calculate the emission from a parsec thick cloud over a grid covering $n_\mathrm{H}=10^4 - 10^{6.5}$ cm$^{-3}$ and $F_X = 1.6-160$ erg\,cm$^{-2}$\,s$^{-1}$. These models generate CO line SEDs with similar shapes as the isothermal models used in our LVG analysis, and an analogous two component fit reproduces the FIR CO line fluxes. In Figure~\ref{coSED}a we show a model that uses $n_\mathrm{H}=10^{5.75}$ cm$^{-3}$ and $F_X=9$ erg\,cm$^{-2}$\,s$^{-1}$ for the ME emission (see Table~\ref{heatingtable}). For an AGN luminosity of $L_\mathrm{2-10\,keV} = 10^{43-44}$ erg\,s$^{-1}$, geometric dilution of the radiation field at the $d\sim100$ pc distance of the H$_2$ ring yields a flux of $F_X=8.4-84$ erg\,cm$^{-2}$\,s$^{-1}$, broadly consistent with this modeled flux. For the plane-parallel geometry employed by~\citet{Meijerink2007} the absolute line luminosities scale with the total XDR surface area, and the normalization of the ME component of the model shown in Figure~\ref{coSED}a requires $A\sim(130$ pc)$^2$. \citet{Galliano2003} model the H$_2$ ring as a section of a 40 pc thick disk. For a radius of 100 pc the inner surface area exposed to the AGN is then $\sim$$(160$ pc)$^2$, similar to the XDR model requirement. In sum, we conclude that if a substantial fraction of the H$_2$ ring is exposed to nuclear hard X-rays, then both the shape of the ME segment of the CO SED and the absolute line fluxes are naturally reproduced in this XDR model.

\begin{table*}
\begin{center}
\caption{Heating Mechanisms\label{heatingtable}}
\begin{tabular}{ccccccccccccrrrrr}
\tableline\tableline
                                  &       ME                           &           HE        &       Full  \\
\tableline
XDR   & $n_\mathrm{H} = 10^{5.75}$ cm$^{-3}$  & $n_\mathrm{H} = 10^{5.25}$ cm$^{-3}$         & ...                 \\
      & $F_X = 9$ erg\,cm$^{-2}$\,s$^{-1}$    & $F_X = 160$ erg\,cm$^{-2}$\,s$^{-1}$         & ...   \\
      & $A\sim(130$ pc)$^2$                   & $A\sim(21$ pc)$^2$  & ...   \\
\tableline
PDR   & ...                                   & $n = 10^{6.5}$ cm$^{-3}$                     & $n = 10^{6}$ cm$^{-3}$               \\
      & ...                                   & $G_0 = 10^{4.75}$                            & $G_0 = 10^{5}$              \\
      & ...                                   & $L_\mathrm{FUV} \sim 2\times10^9$ $L_\odot$  & $L_\mathrm{FUV} \sim 10^{10}$ $L_\odot$   \\
\tableline
shock & C-shock                               &     C-shock     & ...    \\
      & $n_0 = 2\times10^{5}$ cm$^{-3}$       & $n_\mathrm{0} = 10^{6}$ cm$^{-3}$ & ...         \\
      & $v=20$ km\,s$^{-1}$                   & $v=40$ km\,s$^{-1}$ & ...                      \\
      & $A\sim(150$ pc)$^2$                   & $A\sim(16$ pc)$^2$  & ...         \\
\tableline
\end{tabular}
\tablecomments{Details for the models used in Figure~\ref{coSED}. XDR and PDR models are from~\citet{Meijerink2007}, ME C-shock model is from~\citet{Flower2010}, and HE C-shock model is from~\citet{Kaufman1996notmaser}.}
\end{center}
\end{table*}

\subsection{Far-UV Heating} \label{mefuvheating}

Far-UV (FUV; 6 eV $< h\nu <$ 13.6 eV) photons offer another means of heating the molecular gas. Photodissociation regions (PDRs) powered by the FUV radiation from OB stars in the Galaxy have indeed been identified as prominent sources of FIR CO emission~\citep[e.g., ][]{Kramer2004CO}, as have the FUV-irradiated cavities of protostellar outflows~\citep{vanKempen2010CO}. However, a comparison with the high-$J$ CO emission from the prototypical starburst galaxy M82 suggests that the CO emission from NGC 1068 is too strong to be powered by stellar FUV. Herschel observations of M82 have detected the submillimeter CO transitions ($J_\mathrm{upper}=4-13$), which trace emission from gas heated by shocks~\citep{Panuzzo2010} and/or in PDRs~\citep{Loenen2010}. For an $L_\mathrm{FIR}\sim3\times10^{10}$ $L_\odot$~\citep{Telesco1980,Joy1987}, the CO(13-12)/FIR ratio integrated over the central $43.4\arcsec$ ($\sim$$821$ pc) is $\sim$$6\times10^{-6}$. The FIR continuum in M82 is produced by the FUV output of young stars, and we interpret this CO(13-12)/FIR ratio as a benchmark estimate of the fraction of FUV radiation that may be converted to FIR CO emission in an extragalactic starburst. The CO(14-13)/FIR ratio we estimate for NGC 1068 is $\sim$$3\times10^{-5}$, a factor of $\sim$$5$ larger than the CO(13-12)/FIR ratio in M82. This discrepancy is further increased by noting that only a minor fraction of the FIR continuum from the CND in NGC 1068 originates in young stars. The nuclear stellar cluster in NGC 1068 has a characteristic age of $200-300$ Myr, and hence $L_\mathrm{bol}/L_\mathrm{K}\lesssim70$~\citep[for a starburst with exponential decay timescale of $\tau_\mathrm{SF}=100$ Myr;][]{Davies2007ApJ}. For the $L_\mathrm{K}$ measured by~\citet{Davies2007ApJ} this corresponds to $L_\mathrm{bol}\lesssim3\times10^9$ $L_\odot$, more than 5 times lower than the $L_\mathrm{FIR}$ measured here. PDRs heated by the output of this cluster would therefore have to be $\sim$$25$ times more efficient in converting the incident FUV radiation to FIR CO emission than the PDRs in the nucleus of M82. \citet{Loenen2010} model the $J_\mathrm{upper}=1-13$ CO emission from M82 with a 3 component PDR model, in which the CO(13-12) emission is almost entirely generated by the highest excitation component. The CO(13-12)/FIR ratio of this component is comparable to the measured CO(14-13)/FIR ratio in NGC 1068, but the total modeled ratio is a factor of $\sim$$14$ lower due to the FIR emission from the lower excitation material. Increasing the net ratio by a factor of $\sim$$25$ would require the total FIR be dominated by the highest excitation component, and hence a significant suppression of the lower excitation PDR emission. We consider that such an extreme repartitioning of the stellar FUV between M82 and NGC 1068 less likely than the XDR scenario discussed in section~\ref{mexrayheating}.

The CO emission is more plausibly attributed to PDRs powered by the AGN FUV radiation, and in many ways this is an attractive option. The intrinsic AGN luminosity in NGC 1068 of $L_\mathrm{FUV}\sim7.4\times10^{43}$ erg\,s$^{-1}$~\citep{Pier1994} is similar to the observed $L_\mathrm{FIR}$, supporting the idea that the FIR continuum is emission from dust grains heated by the AGN. The measured 60 $\mu$m/100 $\mu$m flux ratio of $\sim$$1.3$ implies that the incident FUV flux on these grains corresponds to $G_0\sim10^5$ (where $F_\mathrm{FUV} = G_0\times1.6\times10^{-3}$ erg\,s$^{-1}$\,cm$^{-2}$)~\citep{Abel2009}, which would also be expected if the AGN $L_\mathrm{FUV}$ is absorbed at the $d\sim100$ pc distance of the H$_2$ ring. To estimate the CO emission produced in these PDRs we employ the type A PDR models of~\citet{Meijerink2007}, which adopt the same geometry and gas densities as the XDR models, and span a FUV intensity range of $G_0 = 10^2 - 10^5$. In dense PDRs most of the CO forms at cloud depths greater than $A_V\approx2$ at gas temperatures $T\lesssim100$ K, but a secondary CO abundance peak at $A_V\approx0.6$ with $T\sim800$ K follows from a local enhancement of OH~\citep{Sternberg1995}. The CO line SEDs generated in the~\citet{Meijerink2007} models are characterized by two distinct peaks, which we associate with these warm and hot CO phases. In Figure~\ref{coSED}b we show a model with $n_\mathrm{H}=10^6$ cm$^{-3}$ and $G_0 = 10^5$ that produces an intriguing match to most (with the exception of CO[30-29]) of the entire FIR CO line SED (solid black line), with the warm and hot phases reproducing the ME and HE components, respectively. Assuming the emitted FIR continuum is equal to the incident FUV, this model predicts a CO(14-13)/FIR ratio of $4\times10^{-5}$. This is similar to the observed ratio in high-$G_0$ Galactic PDRs~\citep{Kramer2004CO}, and the value measured here for NGC 1068.

However, there are at least 2 reasons to reject this scenario. First, the AGN UV/optical emission escapes to $d\sim100$ pc distances only in the ionization cone~\citep[PA $\approx15\arcdeg$][]{Macchetto1994}, while the brightest emission from the H$_2$ ring is to the east. If the warm molecular gas traced by the ME CO emission is assumed to be cospatial with the hot gas traced by the H$_2$ 1-0 S(1) line, then we would expect little interaction between this warm gas and the nuclear FUV. Additionally, any single PDR model that simultaneously matches the ME and HE emission is inconsistent with the kinematic evidence that these two sets of lines are tracing physically distinct components. In general, we require that any model reproducing the $J_\mathrm{upper}\lesssim17$ lines must underpredict the fluxes in the $J_\mathrm{upper}\gsim20$ transitions. In contrast, all of the models produced by~\citet{Meijerink2007} that fit the ME section of the CO SED either match or exceed the observed HE lines fluxes. Consequently we exclude PDRs as a potential source of the ME emission. We stress that this latter argument depends sensitively on an accurate modeling of the $J_\mathrm{upper}\gsim20$ CO emission from the hot surfaces of PDRs. Future Herschel-PACS studies of the FIR CO line SEDs in Galactic PDR templates will be useful in further evaluating this PDR scenario.

\subsection{Shock Heating} \label{meshocks}

Shock heating offers a simple means of exciting the high-$J$ CO emission, and is generally the preferred mechanism for producing the highest excitation molecular gas in Galactic sources~\citep[e.g.,][]{Sempere2000}. Using the C-shock models of~\citet{Flower2010}, we find that the ME component of the CO SED can be fit with a preshock density of $n_\mathrm{H}=2\times10^5$ cm$^{-3}$ and shock velocities of $v_\mathrm{s}=10-20$ km\,s$^{-1}$. The contributions of H$_2$O and H$_2$ to the total cooling budget increase rapidly with shock velocity in these models, and keeping $v_\mathrm{s}\le20$ km\,s$^{-1}$ is necessary to prevent the predicted H$_2$O and H$_2$ line fluxes from exceeding the measured values. Decreasing the shock velocity from $v_\mathrm{s}=20\rightarrow10$ km\,s$^{-1}$ generates weaker CO lines, and requires increasing the total cross-section from $A\sim$ (150 pc)$^2\rightarrow$ (210 pc)$^2$ to match the absolute line fluxes. In Figure~\ref{coSED}c we combine the $v_\mathrm{s}=20$ km\,s$^{-1}$ model with a separate C-shock fit to the HE CO emission.

What shock mechanism could produce the ME CO line emission? Mechanical heating from stellar feedback has been proposed as the energy source behind the submillimeter CO emission in the M82 starburst~\citep{Panuzzo2010} and in other galactic nuclei~\citep{HaileyDunsheath2008,Nikola2011}, but the contrast in CO/FIR ratios between NGC 1068 and M82 discussed in section~\ref{mefuvheating} argues against a similar mechanism here. Furthermore, we can use the results of \citet{Davies2007ApJ} to estimate the mechanical power injected into the ISM by supernovae (SN) in the nuclear cluster of NGC 1068. These authors model the supernovae rate (SNR) to $L_K$ ratio as $\lesssim6\times10^{-11}$ yr$^{-1}$ $L_\odot^{-1}$ in this cluster, which combined with their measured $L_K$ yields SNR $\lesssim2.4\times10^{-3}$. Following~\citet{Loenen2008} and estimating a mechanical energy release of $10^{51}$ erg per SN, with 10\% dissipated in molecular gas, yields a heating rate of $\lesssim$$2\times10^6$ $L_\odot$. This is at least a factor of $\sim$$9-37$ times lower than the mechanical luminosity $L=1/2\rho v_\mathrm{s}^3 A$ required by the shock models discussed above. Jet-ISM interactions are another potential source of shock heating in Seyfert nuclei, but as discussed in section~\ref{linekinematics}, the mismatch in line profiles between the FIR CO lines and the disturbed H$_2$ 1-0 S(1) profile $\sim$$1\arcsec$ north of the AGN suggests jet-driven shocks in the H$_2$ ring may not be important. 

Alternatively, we note that the cross-sections required to normalize these plane-parallel shock models are similar to the estimated cross section of the H$_2$ ring as viewed from the center. The kinematics drawn from the high resolution H$_2$ 1-0 S(1) and millimeter-wave molecular gas maps require a radial expansion component to the H$_2$ ring~\citep{Galliano2002,Davies2008conference,Krips2011}. \citet{Galliano2002} model the H$_2$ dynamics by combining a rotational component with a $v=140$ km\,s$^{-1}$ radially expanding component that generates 1/3 of the total line emission, and \citet{Krips2011} construct a similar model to explain the CO dynamics. We suggest that the interaction of the outflowing gas with non-outflowing material in the H$_2$ ring offers the most plausible source of shock heating. We recall that our excitation model requires the dense gas associated with the ME CO emission to be mixed with lower density (${n_H}_2\lesssim10^5$ cm$^{-3}$) material responsible for the millimeter-wave CO emission (section~\ref{physicalparameters}). Assuming the $nv^2$ product is conserved for shocks propagating through an inhomogeneous medium~\citep{Klein1994}, the moderate ($v\lesssim20$ km\,s$^{-1}$) velocities we require could ultimately be produced in $v\sim140$ km\,s$^{-1}$ shocks triggered in lower density gas. 

\subsection{Southern Streamer} \label{southernheating}

In section~\ref{linekinematics} we showed that the ME CO line profile may be consistent with that of the H$_2$ 1-0 S(1) emission from the southern streamer. However, such an association is inconsistent with either the X-ray or shock heating scenarios outlined above. In both of these models the required cross-section of $A\sim(130$ pc)$^2-(150$ pc)$^2$ is significantly larger than the $\sim$$20$ pc size of the southern streamer. Equivalently, the absolute intensity of the CO emission from this cloud would have to be more than an order of magnitude larger than predicted by the models. As such, we argue that the ME CO emission does not arise from the southern streamer, but most likely arises from the H$_2$ ring.

\subsection{Summary and Discussion} \label{mesummary}

In summary, we conclude that the ME CO emission arises from either X-ray or shock-heated gas in the H$_2$ ring. The challenge of unambiguously determining the heat source for this gas is similar to the situation for the warm and hot gas traced by the H$_2$ rotational and ro-vibrational lines in NGC 1068~\citep{Rotaciuc1991,Lutz2000all}, and in larger samples of Seyferts~\citep{Davies2005,Rodriguez_Ardila2005,Roussel2007}. The link between the ME CO and the H$_2$ emission in NGC 1068 is not immediately clear, although at least to within the PACS resolution the similarity of the ME CO and H$_2$ 1-0 S(1) line profiles suggests a connection (Figure~\ref{lineprofiles}). Additionally, we note that the H$_2$ 1-0 S(1) brightness is quantitatively reproduced with a similar XDR model as used here for the ME CO~\citep[although using a lower density of $n=10^5$ cm$^{-3}$;][]{Maloney1997,Galliano2003}, while the shock model shown in Figure~\ref{coSED}c also produces an H$_2$ 1-0 S(1) flux within a factor of $\sim$$2$ of the total CND emission. Further joint modeling of the CO, H$_2$, and other molecular emission in NGC 1068, and FIR CO data on a larger sample of comparison sources, will be useful in better understanding the nature of the ME CO component.

Here we offer two reasons for preferring the X-ray heating scenario. First, the hard X-ray luminosity of NGC 1068 is reasonably well established through either an analysis of the directly observed (scattered) emission, or by applying scaling relations established for type 1 systems~\citep{Iwasawa1997,Colbert2002}. As discussed above, combining this luminosity with reasonable estimates of the gas density and the CND geometry naturally produces the ME CO line SED. In contrast, it is not clear whether the shocks in the CND are sufficient to dissipate enough mechanical energy at the low velocities needed to power the CO. As we argued above, jet-driven shocks in the H$_2$ ring do not provide a good fit, while the energetics of the possible shocks arising from the radial expansion of the H$_2$ ring have yet to be demonstrated. Secondly, a number of authors have noted that the chemical composition of the CND is best described through a model of X-ray-driven chemistry~\citep{Usero2004,Krips2008,GarciaBurillo2010,Krips2011}. The OH$^+$ and H$_2$O$^+$ emission detected in our PACS scans offers further evidence~\citep{vanderWerf2010,Rangwala2011}, and will be discussed in a future paper. If the nuclear X-rays are responsible for the anomalous molecular abundances in the CND, it is likely they also play an important role in the energetics.

\section{Heating the HE Component} \label{heheating}

The HE CO emission arises from a small mass ($M_{H_2}\sim10^{5.6}$ $M_\odot$) of warm ($T_\mathrm{kin}\sim571$ K) and dense ($n_{H_2}\sim10^{6.4}$ cm$^{-3}$) material that represents only a minor fraction ($\lesssim10\%$) of the total gas in the CND. The kinematic analysis presented in section~\ref{linekinematics} suggests that this component may potentially be associated with the most blueshifted emission in the east or west of the H$_2$ ring, or with the clump of infalling gas $\sim$$0.4\arcsec$ north of the AGN. Here we consider potential heating mechanisms, and argue that the HE CO emission arises from either X-ray-heated gas in the northern streamer, or shock-heated material in either the northern streamer or H$_2$ ring.

\subsection{X-ray Heating} \label{hexrayheating}

In section~\ref{mexrayheating} we used the~\citet{Meijerink2007} XDR models to generate a two component fit to the FIR CO emission, and overploted a sample solution in Figure~\ref{coSED}a. For the $n_\mathrm{H}=10^5 \rightarrow 10^{6.5}$ cm$^{-3}$ density range the ME component requires $F_X = 5.1 \rightarrow 16$ erg\,cm$^{-2}$\,s$^{-1}$, while for the same densities the HE component requires the maximum (or higher) $F_X = 160$ erg\,cm$^{-2}$\,s$^{-1}$ used in the~\citet{Meijerink2007} model grid. An important outcome of this XDR modeling is therefore that the HE component requires irradiation by an X-ray field at least an order of magnitude stronger than the ME component. This is difficult to achieve if both components are presumed to arise from the H$_2$ ring. The eastern segment of the ring lies no more than a factor of $\sim$$1.5$ closer to the AGN than the western segment, so a variation in the distance to the AGN across the ring appears unlikely to produce a factor of $\sim$$10$ variation in the radiation field strength. \citet{GarciaBurillo2010} have used the $6-8$ keV emission detected with Chandra to trace the penetration of nuclear hard X-rays into the cold neutral ISM. Their image suggests a relatively isotropic illumination of the central few hundred parsecs~\citep[see also][]{Ogle2003}. Specifically, while the hard X-rays may be expected to escape to $\sim$$100$ pc scales more easily in the ionization cone, the $6-8$ keV band image is not significantly enhanced along this direction within the H$_2$ ring.

The most straightforward way of generating a higher X-ray intensity is to attribute the HE emission to the northern streamer. In projection, the northern streamer lies a factor of $\sim$$3$ closer to the AGN than does the H$_2$ ring~\citep[$\sim$$0.4\arcsec$ vs $\sim$$1.2\arcsec$;][]{MuellerSanchez2009}, and is therefore expected to see a $\sim$$9$ times stronger $F_X$. A fit to the HE component with $n_\mathrm{H}=10^{5.25}$ cm$^{-3}$ and $F_X = 160$ erg\,cm$^{-2}$\,s$^{-1}$ (Figure~\ref{coSED}a) requires a surface area of $\sim$$(21$ pc$)^2$, close to the $\sim$$14$ pc ($\sim$$0.2\arcsec$) projected lateral size of this clump. We therefore argue that if X-ray heating accounts for both the ME and HE components, the HE component is most plausibly associated with this infalling gas.

\subsection{Far-UV Heating} \label{hefuvheating}

Using the~\citet{Meijerink2007} PDR models we find that for a proper choice of high density ($n_\mathrm{H}\gsim10^6$ cm$^{-3}$) and strong FUV field ($G_0\gsim10^4$), the models reproduce the $J_\mathrm{upper} \approx 19-24$ lines while underpredicting the lower-$J$ fluxes. An example model with $n_\mathrm{H} = 10^{6.5}$ cm$^{-3}$ and $G_0 = 10^{4.75}$ is shown in Figure~\ref{coSED}b (thin blue line). The FUV continuum associated with this component is $L_\mathrm{FUV}\sim2\times10^9$ $L_\odot$, a factor of $\sim$$10$ less than the observed $L_\mathrm{FIR}$. Thus we may consider a scenario in which the ME CO emission arises from X-ray or shock-heated gas in the H$_2$ ring, while the bulk of the HE emission originates in a trace amount of PDR material that makes only a minor contribution to the total continuum emission. These PDR models all fail to reproduce the CO(30-29) transition, however, which in this scenario must therefore arise from a third component of yet more highly excited gas.

\begin{figure}
\epsscale{1.1}
\plotone{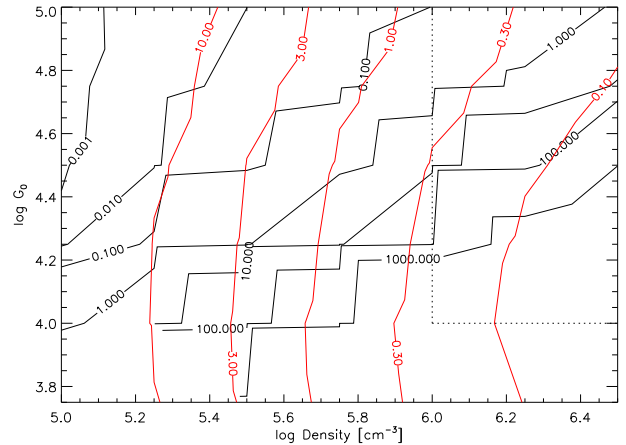}
\caption{Ratio of CO(19-18) emission from a PDR to an XDR (\textit{black}), using the models of~\citet{Meijerink2007}. The incident X-ray flux in the XDR model is 6 times weaker than the FUV flux in the PDR model. The fraction of the measured L$_\mathrm{FIR}$ that must be attributed to this PDR is shown in red. The region of high density and high-$G_0$ that can match the HE CO line SED (excluding CO[30-29]) without overpredicting the lower-$J$ fluxes is indicated with the dotted lines. \label{pdrxdr}}
\end{figure}

Any gas in the CND exposed to the FUV from the AGN must also be irradiated by hard X-rays. For a FUV origin of the HE CO lines the physical properties and radiative environment of the clouds must therefore be such that the FUV is more effective than the X-rays in generating $J_\mathrm{upper}\gsim19$ CO emission. Under what conditions would this occur? A first approach to addressing this question is to separately consider the CO emission from a PDR with that of a properly selected XDR. In Figure~\ref{pdrxdr} we compare the CO(19-18) (the highest-$J$ transition calculated over the full model grid) flux from the~\citet{Meijerink2007} PDR models with that of a~\citet{Meijerink2007} model XDR with $\approx$$1/6$ of the incident flux. This flux ratio is chosen to match the intrinsic $L_\mathrm{FUV}/L_X\approx6$ ratio of the AGN in NGC 1068~\citep{Pier1994}, and is therefore appropriate for a cloud seeing the unattenuated AGN emission. The red curves in Figure~\ref{pdrxdr} show the fraction of the observed FIR continuum (assuming $L_\mathrm{FUV} = L_\mathrm{FIR}$) that must arise from the model PDR in order to account for the absolute CO(20-19) line flux. The region in the upper right quadrant enclosed by the dotted lines indicates the subset of high excitation PDR models that reproduce the HE CO lines (excluding CO[30-29]) while underpredicting the lower-$J$ fluxes -- a requirement for this solution given the velocity shifts between the ME and HE lines. For clouds with $n_\mathrm{H}\gsim10^6$ cm$^{-3}$, $G_0\gsim10^4$, and $G_0/n_\mathrm{H} \lesssim 10^{-1.5}$ cm$^{3}$, this approach suggests that the FUV is more important than the X-rays in generating $J_\mathrm{upper}\gsim19$ CO emission, while only a minor fraction of the AGN FUV luminosity would be required to power such a PDR. 

Where in the nuclear region might these conditions be satisfied? The UV/optical emission from the AGN escapes to large distances within the ionization cone, which runs roughly north-south in projection. If the HE CO emission arises from the H$_2$ ring, however, the line blueshifts would suggest an origin to the east or west (section~\ref{linekinematics}), where little nuclear FUV penetrates. The northern streamer may have a direct view of the AGN, which at a distance of $d\approx40$ pc would correspond to $G_0\sim10^{5.4}$. However, the covering factor of this material (for a size of $\sim$$14$ pc) is only $\sim$$0.01$. The contents of Figure~\ref{pdrxdr} indicate that achieving the absolute HE CO line fluxes with such a small covering factor would require densities far larger than the ${n_H}_2\sim10^{6.4}$ cm$^{-3}$ indicated by our LVG modeling (section~\ref{lvgmodel}). We therefore argue that the HE CO emission is unlikely to be FUV-powered, although more detailed modeling of the conversion of FUV radiation to high-$J$ CO at high densities ($n_\mathrm{H}>10^{6.5}$ cm$^{-3}$) and FUV intensities ($G_0 > 10^5$), and in the presence of an additional X-ray field, would be useful for a more quantitative evaluation of this scenario.

\subsection{Shock Heating}

The HE CO component is fit by a C-shock model with preshock density ${n_H}_2=10^6$ cm$^{-3}$, velocity $v_\mathrm{s}=40$ km\,s$^{-1}$, and cross-section $A\sim$ (16 pc)$^2$~\citep{Kaufman1996notmaser} (Figure~\ref{coSED}c). As discussed in sections~\ref{linekinematics} and~\ref{meshocks}, jet-ISM interactions in the H$_2$ ring are unlikely to produce the FIR CO emission, but a jet-induced shock in the northern streamer is a more plausible source of the HE CO lines. The radio jet changes direction in the vicinity of the brightest clump in the northern streamer, evidence that the fast moving, low density material in the jet is colliding with and being diverted by the dense molecular gas~\citep{MuellerSanchez2009}. This interaction should produce shocks in the molecular material, and the presence of H$_2$O masers in this cloud supports this picture~\citep{Gallimore2004}. The modeled $A\sim$ (16 pc)$^2$ cross-section matches the $\sim$$14$ pc size of this clump, consistent with this scenario. The jet mechanical power estimated in the bow shock model of~\citet{Wilson1987} is$~\sim$$2\times10^8$ $L_\odot$, a factor of $\sim$$3$ larger than the mechanical luminosity $L=1/2\rho v_\mathrm{s}^3 A$ of the shock model discussed here. Jet-induced shocks are therefore energetically feasible, although would require an efficient mechanism for converting kinetic energy at high velocities into slow molecular shocks. Alternatively, shocks in the bluest regions of the H$_2$ ring, possibly associated with the ring expansion, could also generate the HE CO.

An important constraint on this shock modeling is the need to match the CO brightness while not overproducing the ro-vibrational H$_2$ emission. For the C-shock model discussed above the predicted H$_2$ 1-0 S(1) flux is a factor of $\sim$$190$ larger than observed in the northern streamer~\citep{MuellerSanchez2009}, and $\sim$$19$ larger than in the bright eastern clump in the H$_2$ ring~\citep{Galliano2002}. In this region of the~\citet{Kaufman1996notmaser} model parameter space the H$_2$/CO intensity ratio is a steep function of shock velocity, while the CO emission is sensitive to both the velocity and preshock density. Finely-tuning $v_\mathrm{s}$ and $n_0$ (going to lower $v_\mathrm{s}$ and higher $n_0$) may therefore yield a solution that reduces the H$_2$ emission while conserving the CO line SED. Extinction of the 2 $\mu$m H$_2$ emission may also be important, particularly in the northern streamer. \citet{MuellerSanchez2009} estimate a column of $N_\mathrm{H}\sim8\times10^{24}$ cm$^{-2}$ in the southern streamer. Assuming a similar value for the northern streamer, and for $A_K/$N$_\mathrm{H}=(2\times10^{22}$ cm$^{-2})^{-1}$, the resulting $A_K\sim400$ is more than sufficient to provide the necessary attenuation in a mixed dust model. A J-shock model for the HE CO would also generate much weaker H$_2$ emission~\citep{Hollenbach1989}, although the required cross-section would then be a factor of $\sim$$9$ higher than for a C-shock, and no longer match the northern streamer size. In sum, a number of scenarios may be envisioned to attribute the HE CO lines to a shock in the H$_2$ ring or the northern streamer, while also satisfying the H$_2$ 1-0 S(1) brightness.

\subsection{Summary and Discussion} \label{hesummary}

We conclude that the HE component most plausibly arises from X-ray or shock heated gas in the northern streamer, or shock heated gas in the H$_2$ ring. For the ME component, we argued in section~\ref{mesummary} that energetic considerations and the ancillary evidence for an XDR chemistry favored X-ray over shock heating, but the situation for the HE component is less clear. The amount of mechanical power available in the jet or other sources is indeed uncertain, but the strong evidence for a jet interaction with the northern streamer suggests that jet-induced shocks in this cloud be given full consideration. Additionally, the OH$^+$ and H$_2$O$^+$ emission lines that pinpoint an XDR chemistry are emitted close to the galaxy systemic velocity, with no detected emission at the blueshifted velocity of the HE CO lines (Figure~\ref{velcentroid}). Thus while nuclear X-rays may dominate the energetics and chemistry in the CND, an origin of the HE CO lines in a small amount of shock-heated gas must also be considered.

\section{Constraints on the Nuclear Obscuration} \label{nondetections}

\subsection{Could the Detected Emission Arise from Compton-Thick Gas near the AGN?}

The hard X-rays emitted by the AGN in NGC 1068 are obscured by a Compton-thick medium, which may have a column density as high as $N_H>10^{25}$ cm$^{-2}$~\citep{Matt19971068}. Such a high column should provide enough X-ray shielding to enable the formation of CO and other molecules, and this gas may be sufficiently warm and dense to excite the FIR CO transitions~\citep{Krolik1989}. Molecular observations of NGC 1068 have identified two possible candidates for this obscuring material within the central $\lesssim20$ pc. Radio observations at 22 GHz have detected a string of H$_2$O masers that appear to trace a thin, rotating disk, centered on the AGN, with inner and outer radii of $\sim$$0.65$ and $\sim$$1.1$ pc, respectively~\citep[][and references therein]{Gallimore2004}. At larger distances, the southern streamer is observed to approach to within $\sim$$10$ pc of the AGN, and may be the outer part of an amorphous, clumpy structure obscuring the nucleus~\citep{MuellerSanchez2009}. Throughout this paper we have attributed the detected FIR CO emission to gas associated with either the H$_2$ ring or the northern streamer, with the implication that we are not detecting the obscuring medium. In sections~\ref{highreskinematics} and~\ref{southernheating} we explicitly argued that the line profiles and/or absolute intensities of the ME and HE components are inconsistent with either component arising from the southern streamer. However, is it possible that the HE emission arises from the maser disk, or elsewhere within the central few parsecs? And could this gas provide the $N_H\sim10^{25}$ cm$^{-2}$ obscuring column?

The geometrical constraints imposed by the LVG modeling do, in fact, allow for the HE component to comprise a parsec-scale, Compton-thick structure. The LVG model used here assumes emission from a collection of spherical clouds, and retains the freedom to arrange these clouds in an arbitrary configuration. As an example, we consider smoothly distributing the gas in a spherical shell with a finite covering factor ($f_\mathrm{cov}$). Hard X-ray surveys conducted with INTEGRAL/\textit{IBIS}~\citep{Malizia2009} and \textit{Swift}-BAT~\citep{Burlon2011} indicate $\sim$$20-25\%$ of AGN are Compton-thick, and in the spirit of unification we therefore adopt $f_\mathrm{cov}=0.25$. We require the shell to have a column density of $N_H=10^{25}$ cm$^{-2}$, in which case a choice of density and total mass determines the inner radius ($R_\mathrm{in}$). For the set of good-fitting LVG solutions discussed in section~\ref{goodfits}, we find values of $R_\mathrm{in} = 0-7$ pc. This range of radii allows structures matched in size to the maser disk, as well as moderately more extended configurations.

The $\sim$$250$ km\,s$^{-1}$ FWHM of the FIR CO lines is considerably lower than the $\sim$$600$ km\,s$^{-1}$ velocity range of the maser spots. If the HE component is attributed to a uniform disk, virial considerations then require a larger size scale than the $r=0.65-1.1$ pc radial extent of the maser disk. Translating the observed linewidth to a disk size is beyond the scope of this paper. However, the rotational curve of the maser disk scales as $v\propto r^{-0.31}$~\citep{Greenhill1996ApJ}, so even a factor of 1.5 decrease in the circular velocity would require a factor of $\sim$$4$ increase in the radial scale, placing the CO-emitting gas well outside of the maser disk. Alternatively, it would be possible to place the HE component within the maser disk if we allow that a non-uniform excitation or mass distribution enhances the CO emission within the narrower observed velocity range.

The strongest argument against associating the HE component with the maser disk follows from a comparison of the physical parameters, as the density and/or thermal pressure we estimate for the HE component is likely to be smaller than that of the disk. A first estimate of the pressure in this region may be obtained by considering the free-free emitting plasma lying just inside the maser disk ($r\approx0.4$ pc), for which~\citet{Gallimore2004} estimate $n_eT_e\sim(6\times10^5$ cm$^{-3})(6\times10^6$ K$)\sim10^{12.6}$ K\,cm$^{-3}$. This is a factor of $\sim$$10^3$ larger than the range $P/k_B\sim10^{8.8}-10^{9.8}$ K\,cm$^{-3}$ derived from our LVG modeling of the HE component (Table~\ref{lvgtable}). Separately, \citet{Lodato20031068} have derived the surface density profile of the maser disk required to reproduce the non-Keplerian rotation curve, and find a density of ${n_H}_2=(1-5)\times10^8$ cm$^{-3}$ at the outer edge of the disk. Interferometric mid-IR observations have identified a structure of hot ($T\sim800$ K) dust with a size of $\sim$$0.45\times1.35$ pc, which may coincide with the maser disk~\citep{Jaffe2004,Raban2009}. This dust temperature sets a lower limit to the temperature of the concomitant gas. Densities larger than $10^8$ cm$^{-3}$ are in excess of the ${n_H}_2\sim10^{5.9}-10^{7.1}$ cm$^{-3}$ range derived from our LVG modeling (Table~\ref{lvgtable}), and models simultaneously requiring ${n_H}_2 > 10^8$ cm$^{-3}$ and $T_\mathrm{kin} > 800$ K are even more firmly excluded (Figure~\ref{chisqrdplot}).

How far from the AGN would it be possible to find molecular gas at the relatively low pressure we attribute to the HE component? \cite{Neufeld1994maser} and~\cite{Neufeld19954258} have investigated the properties of dense gas in close proximity to a strong X-ray source, and find that such gas is molecular if the pressure exceeds
\begin{equation}
P/k_B\gsim10^{11}L_{43.5}R_\mathrm{pc}^{-2}N_{24}^{-0.9} \mathrm{\,\,K\,cm}^{-3}, 
\label{torpressure}
\end{equation}
\noindent where $10^{43.5}L_{43.5}$ erg\,s$^{-1}$ is the $2-10$ keV luminosity, the gas is $R_\mathrm{pc}$ pc from the source, and is shielded by a column of $N_H=10^{24}N_{24}$ cm$^{-2}$. For NGC 1068 we estimate $L_{43.5}=1$ (see section~\ref{upperlimits}), and for the maser disk we also set $R_\mathrm{pc}=1$. Assuming the maser disk is not shielded by Compton-thick material inward of 1 pc (i.e., setting $N_{24} \le 1$), this expression supports our previous conclusion that the molecular gas in the maser disk is at higher pressure than our HE component. If the HE emission traces the Compton-thick structure blocking the hard X-rays, then by construction any material between this gas and the AGN must be Compton-thin ($N_{24} \lesssim 1$). For $P/k_B\sim10^{8.8}-10^{9.8}$, equation~\ref{torpressure} then places this gas at a distance of $R_\mathrm{pc}\sim4-13$. 

We conclude that the HE CO emission does not arise from the maser disk, primarily due to the large difference in thermal pressures. However, the observed linewidths and the results from our LVG modeling do allow us to construct a model in which the HE component traces the nuclear obscuring material, provided this material lies at least a few parsecs from the AGN. This scale may be broadly consistent with clumpy torus models, in which the obscuring medium is comprised of clouds distributed from the sublimation radius ($r_\mathrm{sub}\lesssim1$ pc) to several parsec scales~\citep[e.g.,][]{Hoenig2006,Nenkova2008II}. These models are currently constrained primarily by IR continuum observations. The inclusion of a gas phase in these models would be a useful next step to further evaluate the prospects for attributing the FIR CO emission to gas in close proximity of the AGN.

\subsection{A Comparison with the~\citet{Krolik1989} Torus Model} \label{upperlimits}

\begin{figure}
\epsscale{1.1}
\plotone{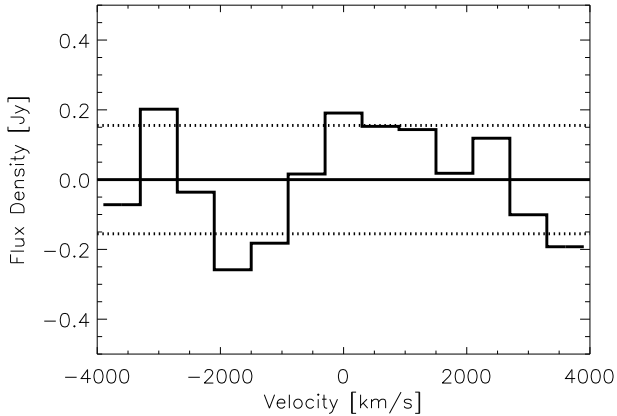}
\caption{Baseline-subtracted stack of 8 transitions with $J_\mathrm{upper}=34-47$, for comparison with the~\citet{Krolik1989} models. Dashed lines mark the $\pm1\sigma$ uncertainty.\label{stacked_KL}}
\end{figure}

In addition to the detected $J_\mathrm{upper}\le30$ transitions, our upper limits to the $J_\mathrm{upper}\le50$ lines provide a useful constraint on any potential high excitation nuclear molecular component. \citet{Krolik1989} modeled the molecular emission expected from a Compton-thick, parsec-scale torus. They calculated the CO emission from an $N_\mathrm{H}=10^{24}$ cm$^{-2}$ cloud located $\sim$$1$ pc from a luminous ($L_X\sim10^{44}$ erg\,s$^{-1}$) hard X-ray source, in which the heating is dominated by Compton scattering of $10-100$ keV photons. In this model the CO SED scales as $J_\mathrm{upper}^3$ up to $J_\mathrm{upper}\approx58$, and the absolute line luminosities are proportional to the total absorbed $10-100$ keV luminosity ($f_\mathrm{abs}L_{x44}$; in units of $10^{44}$ erg\,s$^{-1}$). Several of our upper limits to CO transitions with $J_\mathrm{upper}=34-47$ independently place an upper limit to this component corresponding to $f_\mathrm{abs}L_{x44} \lesssim 0.1$, while the non-detection of the CO(44-43) line achieves the most stringent limit of $f_\mathrm{abs}L_{x44} \lesssim 0.09$ (Figure~\ref{coSED}a). This limit may be reduced by stacking the individual non-detections. We have obtained such a stack by first scaling the spectrum of each undetected line by ($44/J_\mathrm{upper}$)$^2$, which references each spectrum to that of CO(44-43) under the assumption that the CO fluxes scale as $J_\mathrm{upper}^3$. We then calculated a weighted average of 8 lines with low noise that fall in clean spectral regions, and binned to 600 km\,s$^{-1}$. The result is shown in Figure~\ref{stacked_KL}. The upper limit to this stacked line pushes the limit of the~\citet{Krolik1989} model to $f_\mathrm{abs}L_{x44} \lesssim 0.038$ (Figure~\ref{coSED}a). 

\citet{Iwasawa1997} and~\citet{Colbert2002} model the reflected X-ray spectrum of NGC 1068, and estimate intrinsic luminosities of $L_\mathrm{2-10\,keV} = 10^{43-43.7}$ erg\,s$^{-1}$ (corrected to the $d=14.4$ Mpc used here). A comparison of the measured [OIII] $\lambda5007$ line luminosity with a log([OIII]/$L_X)=-1.89$ ratio established for type 1 Seyferts yields $L_\mathrm{2-10\,keV} = 10^{43.5}$ erg\,s$^{-1}$, while a comparison of the estimated $L_\mathrm{bol}$ with $L_\mathrm{2-10\,keV}/L_\mathrm{bol}\sim0.1$ yields $L_\mathrm{2-10\,keV} \sim 10^{43.6}$ erg\,s$^{-1}$~\citep[][and references therein]{Colbert2002}. \citet{Melendez2008} derive a relation between $L_\mathrm{[OIV]\,26\mu m}$ and $L_\mathrm{2-10\,keV}$ for a sample of hard (14-195 keV) X-ray selected type 1 Seyferts. Using this relation, along with $L_\mathrm{[OIV]\,26\mu m} = 4.7\times10^{41}$ erg\,s$^{-1}$~\citep{Sturm2002}, yields $L_\mathrm{2-10\,keV} \approx 10^{44}$ erg\,s$^{-1}$. Given this range of estimates we adopt $L_\mathrm{2-10\,keV} = 10^{43.5}$ erg\,s$^{-1}$, and increase this by $1.43$ to correct to the $10-100$ keV range (for $L_\nu \propto \nu^{-1}$). Assuming 25\% of this power (for a covering fraction $f_\mathrm{cov}=0.25$) is absorbed in the phase modeled by~\citet{Krolik1989} gives an expected $f_\mathrm{abs}L_{x44} = 0.11$, a modest factor of $\sim$$2.9$ higher than our measured upper limit. Given the uncertainties in the AGN luminosity and covering factor of the Comption-thick material in NGC 1068, as well as uncertainties in the model itself, this factor of $\sim$$2.9$ overprediction is too small to reject the model of a parsec-scale torus envisioned by~\citet{Krolik1989}. This non-detection may, however, provide a useful constraint on the more recent set of detailed clumpy and dynamical torus models~\citep[e.g.,][]{Hoenig2006,Nenkova2008II,Wada2009AGN,Schartmann2010}. 

\section{Summary and Future Work} \label{conclusions}

We have detected 11 CO transitions in the $J_\mathrm{upper}=14-30$ range from the central $10\arcsec$ ($700$ pc) of NGC 1068, and obtained sensitive upper limits to most other transitions up to $J_\mathrm{upper}\le50$. These are the first extragalactic detections of FIR CO, which represent a new probe of excited molecular gas in Seyfert nuclei. The detected transitions are modeled as arising from 2 different components: a moderate excitation (ME) component close to the galaxy systemic velocity, and a high excitation (HE) component that is blueshifted by $\sim$$80$ km\,s$^{-1}$. Our main results are as follows:

1) Using a two component LVG model we derive ${n_\mathrm{H}}_2\sim10^{5.6}$ cm$^{-3}$, $T_\mathrm{kin}\sim170$ K, and ${M_H}_2\sim10^{6.7}$ $M_\odot$ for the ME component, and ${n_\mathrm{H}}_2\sim10^{6.4}$ cm$^{-3}$, $T_\mathrm{kin}\sim570$ K, and ${M_H}_2\sim10^{5.6}$ $M_\odot$ for the HE component. The $1\sigma$ uncertainties on the derived temperatures are $\pm(0.20-0.35)$ dex, while for density and mass this is $\pm(0.6-0.9)$ dex. Extending the measured CO line SED to lower- and higher-$J$ lines would likely reduce these uncertainties, as would a joint analysis with H$_2$, OH, H$_2$O, and other complementary molecular tracers. We will follow both approaches in forthcoming papers.

2) These two components are denser than the gas traced with millimeter CO observations, and the HE (and possibly the ME) component is also warmer. The ME component makes a non-negligible contribution to the nuclear mass budget, although large uncertainties in the masses estimated from both the FIR CO lines and from CO(1-0) prevent a more quantitative statement.

3) Comparing the FIR CO line profiles with those of high spatial and spectral resolution observations of CO(2-1) and H$_2$ 1-0 S(1) allows a first estimate of the origins of the ME and HE components within the central $10\arcsec$. Good matches are found with H$_2$ 1-0 S(1), which for the ME component suggests an origin in the $\sim$$200$ pc diameter H$_2$ ring. The blueshifted HE lines may also be consistent with an origin in the bluest regions of the H$_2$ ring, but are better matched to the clump of infalling molecular gas $\sim$$40$ pc north of the AGN.

4) The ME component is nicely consistent with models of X-ray heating of gas in the CND. A shock model is also possible, although due to the uncertainties in the amount of mechanical power available for dissipation in slow shocks, and the evidence for X-ray driven chemistry in the CND, we favor the X-ray heating scenario. Far-UV heating is unlikely.

5) The HE component is also consistent with either X-ray or shock heating. X-ray heating would best fit with an origin in the cloud $\sim$$40$ pc north of the AGN, supporting the results of the line profile matching (point 3). Shocks triggered by the interaction of the radio jet with this clump, or arising from the H$_2$ ring, are also plausible. Far-UV heating is unlikely.

6) The thermal pressure of our HE component is too low to be attributed to gas within the parsec-scale H$_2$O maser disk centered on the AGN. However, the pressure may be consistent with gas located $\sim$$4$ pc or more from the AGN, and this gas could potentially provide the $N_H\sim10^{25}$ cm$^{-2}$ column obscuring the nuclear hard X-rays. Our non-detections of $J_\mathrm{upper}=34-47$ lines place an upper limit to the~\citet{Krolik1989} torus model that is a factor of $\sim$$2.9$ lower than the expected signal, although due to the uncertainties involved in applying this model, this non-detection is insufficient to rule out the parsec-scale torus paradigm. The inclusion of a gas phase in the current set of clumpy and dynamic torus models, and a comparison of the predicted CO line SED with our detections and upper limits, would be a useful next step.

\acknowledgments

We thank Eva Schinnerer for generously providing her IRAM PdBI CO(1-0) and CO(2-1) data, and Francisco M{\"u}ller S{\'a}nchez for providing his SINFONI/VLT H$_2$ 1-0 S(1) map. We also thank an anonymous referee for many helpful comments on an early draft of this manuscript. Basic research in IR astronomy at NRL is funded by the US ONR; J.F. also acknowledges support from the NHSC. We thank the DFG for support via German-Israel Project Cooperation grant STE1869/1-1.GE625/15-1. E.G-A thanks the support by the Spanish Ministerio de Ciencia e Innovaci\'on under project AYA2010-21697-C05-01, and is a Research Associate at the Harvard-Smithsonian Center for Astrophysics. S.V. acknowledges support from a Senior NPP Award from NASA and thanks the host institution, the Goddard Space Flight Center. PACS has been developed by a consortium of institutes led by MPE (Germany) and including UVIE (Austria); KU Leuven, CSL, IMEC (Belgium); CEA, LAM (France); MPIA (Germany); INAF-IFSI/OAA/OAP/OAT, LENS, SISSA (Italy); IAC (Spain). This development has been supported by the funding agencies BMVIT (Austria), ESA-PRODEX (Belgium), CEA/ CNES (France), DLR (Germany), ASI/ INAF (Italy), and CICYT/ MCYT (Spain).


\begin{thebibliography}{95}
\expandafter\ifx\csname natexlab\endcsname\relax\def\natexlab#1{#1}\fi

\bibitem[{{Aalto} {et~al.}(2011){Aalto}, {Costagliola}, {van der Tak}, \&
  {Meijerink}}]{Aalto2011}
{Aalto}, S., {Costagliola}, F., {van der Tak}, F., \& {Meijerink}, R. 2011,
  \aap, 527, A69

\bibitem[{{Abel} {et~al.}(2009){Abel}, {Dudley}, {Fischer}, {Satyapal}, \& {van
  Hoof}}]{Abel2009}
{Abel}, N.~P., {Dudley}, C., {Fischer}, J., {Satyapal}, S., \& {van Hoof},
  P.~A.~M. 2009, \apj, 701, 1147

\bibitem[{{Axon} {et~al.}(1998){Axon}, {Marconi}, {Capetti}, {Maccetto},
  {Schreier}, \& {Robinson}}]{Axon1998}
{Axon}, D.~J., {Marconi}, A., {Capetti}, A., {Maccetto}, F.~D., {Schreier}, E.,
  \& {Robinson}, A. 1998, \apjl, 496, L75

\bibitem[{{Bland-Hawthorn} {et~al.}(1997){Bland-Hawthorn}, {Gallimore},
  {Tacconi}, {Brinks}, {Baum}, {Antonucci}, \& {Cecil}}]{Bland_Hawthorn1997}
{Bland-Hawthorn}, J., {Gallimore}, J.~F., {Tacconi}, L.~J., {Brinks}, E.,
  {Baum}, S.~A., {Antonucci}, R.~R.~J., \& {Cecil}, G.~N. 1997, \apss, 248, 9

\bibitem[{{Blietz} {et~al.}(1994){Blietz}, {Cameron}, {Drapatz}, {Genzel},
  {Krabbe}, {van der Werf}, {Sternberg}, \& {Ward}}]{Blietz1994}
{Blietz}, M., {Cameron}, M., {Drapatz}, S., {Genzel}, R., {Krabbe}, A., {van
  der Werf}, P., {Sternberg}, A., \& {Ward}, M. 1994, \apj, 421, 92

\bibitem[{{Bryant} \& {Scoville}(1996)}]{Bryant1996}
{Bryant}, P.~M., \& {Scoville}, N.~Z. 1996, \apj, 457, 678

\bibitem[{{Burlon} {et~al.}(2011){Burlon}, {Ajello}, {Greiner}, {Comastri},
  {Merloni}, \& {Gehrels}}]{Burlon2011}
{Burlon}, D., {Ajello}, M., {Greiner}, J., {Comastri}, A., {Merloni}, A., \&
  {Gehrels}, N. 2011, \apj, 728, 58

\bibitem[{{Cameron} {et~al.}(1993){Cameron}, {Storey}, {Rotaciuc}, {Genzel},
  {Verstraete}, {Drapatz}, {Siebenmorgen}, \& {Lee}}]{Cameron1993}
{Cameron}, M., {Storey}, J.~W.~V., {Rotaciuc}, V., {Genzel}, R., {Verstraete},
  L., {Drapatz}, S., {Siebenmorgen}, R., \& {Lee}, T.~J. 1993, \apj, 419, 136

\bibitem[{{Colbert} {et~al.}(2002){Colbert}, {Weaver}, {Krolik}, {Mulchaey}, \&
  {Mushotzky}}]{Colbert2002}
{Colbert}, E.~J.~M., {Weaver}, K.~A., {Krolik}, J.~H., {Mulchaey}, J.~S., \&
  {Mushotzky}, R.~F. 2002, \apj, 581, 182

\bibitem[{{Davies} {et~al.}(2008){Davies}, {Genzel}, {Tacconi}, {S{\'a}nchez},
  \& {Sternberg}}]{Davies2008conference}
{Davies}, R., {Genzel}, R., {Tacconi}, L., {S{\'a}nchez}, F.~M., \&
  {Sternberg}, A. 2008, in Mapping the Galaxy and Nearby Galaxies, ed. {K.~Wada
  \& F.~Combes}, 144--+

\bibitem[{{Davies} {et~al.}(2007){Davies}, {S{\'a}nchez}, {Genzel}, {Tacconi},
  {Hicks}, {Friedrich}, \& {Sternberg}}]{Davies2007ApJ}
{Davies}, R.~I., {S{\'a}nchez}, F.~M., {Genzel}, R., {Tacconi}, L.~J., {Hicks},
  E.~K.~S., {Friedrich}, S., \& {Sternberg}, A. 2007, \apj, 671, 1388

\bibitem[{{Davies} {et~al.}(2005){Davies}, {Sternberg}, {Lehnert}, \&
  {Tacconi-Garman}}]{Davies2005}
{Davies}, R.~I., {Sternberg}, A., {Lehnert}, M.~D., \& {Tacconi-Garman}, L.~E.
  2005, \apj, 633, 105

\bibitem[{{Davies} {et~al.}(1998){Davies}, {Sugai}, \& {Ward}}]{Davies1998}
{Davies}, R.~I., {Sugai}, H., \& {Ward}, M.~J. 1998, \mnras, 300, 388

\bibitem[{{Flower} \& {Pineau Des For{\^e}ts}(2010)}]{Flower2010}
{Flower}, D.~R., \& {Pineau Des For{\^e}ts}, G. 2010, \mnras, 406, 1745

\bibitem[{{Frerking} {et~al.}(1982){Frerking}, {Langer}, \&
  {Wilson}}]{Frerking1982}
{Frerking}, M.~A., {Langer}, W.~D., \& {Wilson}, R.~W. 1982, \apj, 262, 590

\bibitem[{{Galliano} \& {Alloin}(2002)}]{Galliano2002}
{Galliano}, E., \& {Alloin}, D. 2002, \aap, 393, 43

\bibitem[{{Galliano} {et~al.}(2003){Galliano}, {Alloin}, {Granato}, \&
  {Villar-Mart{\'{\i}}n}}]{Galliano2003}
{Galliano}, E., {Alloin}, D., {Granato}, G.~L., \& {Villar-Mart{\'{\i}}n}, M.
  2003, \aap, 412, 615

\bibitem[{{Gallimore} {et~al.}(2004){Gallimore}, {Baum}, \&
  {O'Dea}}]{Gallimore2004}
{Gallimore}, J.~F., {Baum}, S.~A., \& {O'Dea}, C.~P. 2004, \apj, 613, 794

\bibitem[{{Gallimore} {et~al.}(1996){Gallimore}, {Baum}, {O'Dea}, \&
  {Pedlar}}]{Gallimore1996paperI}
{Gallimore}, J.~F., {Baum}, S.~A., {O'Dea}, C.~P., \& {Pedlar}, A. 1996, \apj,
  458, 136

\bibitem[{{Gallimore} {et~al.}(2001){Gallimore}, {Henkel}, {Baum}, {Glass},
  {Claussen}, {Prieto}, \& {Von Kap-herr}}]{Gallimore2001}
{Gallimore}, J.~F., {Henkel}, C., {Baum}, S.~A., {Glass}, I.~S., {Claussen},
  M.~J., {Prieto}, M.~A., \& {Von Kap-herr}, A. 2001, \apj, 556, 694

\bibitem[{{Garc{\'{\i}}a-Burillo} {et~al.}(2010){Garc{\'{\i}}a-Burillo},
  {Usero}, {Fuente}, {Mart{\'{\i}}n-Pintado}, {Boone}, {Aalto}, {Krips},
  {Neri}, {Schinnerer}, \& {Tacconi}}]{GarciaBurillo2010}
{Garc{\'{\i}}a-Burillo}, S. {et~al.} 2010, \aap, 519, A2+

\bibitem[{{Goldsmith}(2001)}]{Goldsmith2001}
{Goldsmith}, P.~F. 2001, \apj, 557, 736

\bibitem[{{Greenhill} {et~al.}(1996){Greenhill}, {Gwinn}, {Antonucci}, \&
  {Barvainis}}]{Greenhill1996ApJ}
{Greenhill}, L.~J., {Gwinn}, C.~R., {Antonucci}, R., \& {Barvainis}, R. 1996,
  \apjl, 472, L21

\bibitem[{{Hailey-Dunsheath} {et~al.}(2008){Hailey-Dunsheath}, {Nikola},
  {Stacey}, {Oberst}, {Parshley}, {Bradford}, {Ade}, \&
  {Tucker}}]{HaileyDunsheath2008}
{Hailey-Dunsheath}, S., {Nikola}, T., {Stacey}, G.~J., {Oberst}, T.~E.,
  {Parshley}, S.~C., {Bradford}, C.~M., {Ade}, P.~A.~R., \& {Tucker}, C.~E.
  2008, \apjl, 689, L109

\bibitem[{{Helfer} \& {Blitz}(1995)}]{Helfer1995}
{Helfer}, T.~T., \& {Blitz}, L. 1995, \apj, 450, 90

\bibitem[{{Hollenbach} \& {McKee}(1989)}]{Hollenbach1989}
{Hollenbach}, D., \& {McKee}, C.~F. 1989, \apj, 342, 306

\bibitem[{{H{\"o}nig} {et~al.}(2006){H{\"o}nig}, {Beckert}, {Ohnaka}, \&
  {Weigelt}}]{Hoenig2006}
{H{\"o}nig}, S.~F., {Beckert}, T., {Ohnaka}, K., \& {Weigelt}, G. 2006, \aap,
  452, 459

\bibitem[{{Israel}(2009)}]{Israel20091068}
{Israel}, F.~P. 2009, \aap, 493, 525

\bibitem[{{Iwasawa} {et~al.}(1997){Iwasawa}, {Fabian}, \& {Matt}}]{Iwasawa1997}
{Iwasawa}, K., {Fabian}, A.~C., \& {Matt}, G. 1997, \mnras, 289, 443

\bibitem[{{Jaffe} {et~al.}(2004){Jaffe}, {Meisenheimer}, {R{\"o}ttgering},
  {Leinert}, {Richichi}, {Chesneau}, {Fraix-Burnet}, {Glazenborg-Kluttig},
  {Granato}, {Graser}, {Heijligers}, {K{\"o}hler}, {Malbet}, {Miley},
  {Paresce}, {Pel}, {Perrin}, {Przygodda}, {Schoeller}, {Sol}, {Waters},
  {Weigelt}, {Woillez}, \& {de Zeeuw}}]{Jaffe2004}
{Jaffe}, W. {et~al.} 2004, \nat, 429, 47

\bibitem[{{Joy} {et~al.}(1987){Joy}, {Lester}, \& {Harvey}}]{Joy1987}
{Joy}, M., {Lester}, D.~F., \& {Harvey}, P.~M. 1987, \apj, 319, 314

\bibitem[{{Kamenetzky} {et~al.}(2011){Kamenetzky}, {Glenn}, {Maloney},
  {Aguirre}, {Bock}, {Bradford}, {Earle}, {Inami}, {Matsuhara}, {Murphy},
  {Naylor}, {Nguyen}, \& {Zmuidzinas}}]{Kamenetzky2011}
{Kamenetzky}, J. {et~al.} 2011, \apj, 731, 83

\bibitem[{{Kaufman} \& {Neufeld}(1996)}]{Kaufman1996notmaser}
{Kaufman}, M.~J., \& {Neufeld}, D.~A. 1996, \apj, 456, 611

\bibitem[{{Klein} {et~al.}(1994){Klein}, {McKee}, \& {Colella}}]{Klein1994}
{Klein}, R.~I., {McKee}, C.~F., \& {Colella}, P. 1994, \apj, 420, 213

\bibitem[{{Kramer} {et~al.}(2004){Kramer}, {Jakob}, {Mookerjea}, {Schneider},
  {Br{\"u}ll}, \& {Stutzki}}]{Kramer2004CO}
{Kramer}, C., {Jakob}, H., {Mookerjea}, B., {Schneider}, N., {Br{\"u}ll}, M.,
  \& {Stutzki}, J. 2004, \aap, 424, 887

\bibitem[{{Krips} {et~al.}(2011){Krips}, {Mart{\'{\i}}n}, {Eckart}, {Neri},
  {Garc{\'{\i}}a-Burillo}, {Matsushita}, {Peck}, {Stoklasov{\'a}}, {Petitpas},
  {Usero}, {Combes}, {Schinnerer}, {Humphreys}, \& {Baker}}]{Krips2011}
{Krips}, M. {et~al.} 2011, \apj, 736, 37

\bibitem[{{Krips} {et~al.}(2008){Krips}, {Neri}, {Garc{\'{\i}}a-Burillo},
  {Mart{\'{\i}}n}, {Combes}, {Graci{\'a}-Carpio}, \& {Eckart}}]{Krips2008}
{Krips}, M., {Neri}, R., {Garc{\'{\i}}a-Burillo}, S., {Mart{\'{\i}}n}, S.,
  {Combes}, F., {Graci{\'a}-Carpio}, J., \& {Eckart}, A. 2008, \apj, 677, 262

\bibitem[{{Krolik} \& {Lepp}(1989)}]{Krolik1989}
{Krolik}, J.~H., \& {Lepp}, S. 1989, \apj, 347, 179

\bibitem[{{Le Floc'h} {et~al.}(2001){Le Floc'h}, {Mirabel}, {Laurent},
  {Charmandaris}, {Gallais}, {Sauvage}, {Vigroux}, \& {Cesarsky}}]{LeFloch2001}
{Le Floc'h}, E., {Mirabel}, I.~F., {Laurent}, O., {Charmandaris}, V.,
  {Gallais}, P., {Sauvage}, M., {Vigroux}, L., \& {Cesarsky}, C. 2001, \aap,
  367, 487

\bibitem[{{Lodato} \& {Bertin}(2003)}]{Lodato20031068}
{Lodato}, G., \& {Bertin}, G. 2003, \aap, 398, 517

\bibitem[{{Loenen} {et~al.}(2008){Loenen}, {Spaans}, {Baan}, \&
  {Meijerink}}]{Loenen2008}
{Loenen}, A.~F., {Spaans}, M., {Baan}, W.~A., \& {Meijerink}, R. 2008, \aap,
  488, L5

\bibitem[{{Loenen} {et~al.}(2010){Loenen}, {van der Werf}, {G{\"u}sten},
  {Meijerink}, {Israel}, {Requena-Torres}, {Garc{\'{\i}}a-Burillo}, {Harris},
  {Klein}, {Kramer}, {Lord}, {Mart{\'{\i}}n-Pintado}, {R{\"o}llig}, {Stutzki},
  {Szczerba}, {Wei{\ss}}, {Philipp-May}, {Yorke}, {Caux}, {Delforge},
  {Helmich}, {Lorenzani}, {Morris}, {Philips}, {Risacher}, \&
  {Tielens}}]{Loenen2010}
{Loenen}, A.~F. {et~al.} 2010, \aap, 521, L2

\bibitem[{{Lutz} {et~al.}(2000){Lutz}, {Sturm}, {Genzel}, {Moorwood},
  {Alexander}, {Netzer}, \& {Sternberg}}]{Lutz2000all}
{Lutz}, D., {Sturm}, E., {Genzel}, R., {Moorwood}, A.~F.~M., {Alexander}, T.,
  {Netzer}, H., \& {Sternberg}, A. 2000, \apj, 536, 697

\bibitem[{{Macchetto} {et~al.}(1994){Macchetto}, {Capetti}, {Sparks}, {Axon},
  \& {Boksenberg}}]{Macchetto1994}
{Macchetto}, F., {Capetti}, A., {Sparks}, W.~B., {Axon}, D.~J., \&
  {Boksenberg}, A. 1994, \apjl, 435, L15

\bibitem[{{Malizia} {et~al.}(2009){Malizia}, {Stephen}, {Bassani}, {Bird},
  {Panessa}, \& {Ubertini}}]{Malizia2009}
{Malizia}, A., {Stephen}, J.~B., {Bassani}, L., {Bird}, A.~J., {Panessa}, F.,
  \& {Ubertini}, P. 2009, \mnras, 399, 944

\bibitem[{{Maloney}(1997)}]{Maloney1997}
{Maloney}, P.~R. 1997, \apss, 248, 105

\bibitem[{{Maloney} {et~al.}(1996){Maloney}, {Hollenbach}, \&
  {Tielens}}]{Maloney1996}
{Maloney}, P.~R., {Hollenbach}, D.~J., \& {Tielens}, A.~G.~G.~M. 1996, \apj,
  466, 561

\bibitem[{{Matt} {et~al.}(1997){Matt}, {Guainazzi}, {Frontera}, {Bassani},
  {Brandt}, {Fabian}, {Fiore}, {Haardt}, {Iwasawa}, {Maiolino}, {Malaguti},
  {Marconi}, {Matteuzzi}, {Molendi}, {Perola}, {Piraino}, \&
  {Piro}}]{Matt19971068}
{Matt}, G. {et~al.} 1997, \aap, 325, L13

\bibitem[{{Meijerink} \& {Spaans}(2005)}]{Meijerink2005}
{Meijerink}, R., \& {Spaans}, M. 2005, \aap, 436, 397

\bibitem[{{Meijerink} {et~al.}(2007){Meijerink}, {Spaans}, \&
  {Israel}}]{Meijerink2007}
{Meijerink}, R., {Spaans}, M., \& {Israel}, F.~P. 2007, \aap, 461, 793

\bibitem[{{Mel{\'e}ndez} {et~al.}(2008){Mel{\'e}ndez}, {Kraemer}, {Armentrout},
  {Deo}, {Crenshaw}, {Schmitt}, {Mushotzky}, {Tueller}, {Markwardt}, \&
  {Winter}}]{Melendez2008}
{Mel{\'e}ndez}, M. {et~al.} 2008, \apj, 682, 94

\bibitem[{{Miller} \& {Antonucci}(1983)}]{Miller1983scatter}
{Miller}, J.~S., \& {Antonucci}, R.~R.~J. 1983, \apjl, 271, L7

\bibitem[{{Mouri}(1994)}]{Mouri1994}
{Mouri}, H. 1994, \apj, 427, 777

\bibitem[{{M{\"u}ller S{\'a}nchez} {et~al.}(2009){M{\"u}ller S{\'a}nchez},
  {Davies}, {Genzel}, {Tacconi}, {Eisenhauer}, {Hicks}, {Friedrich}, \&
  {Sternberg}}]{MuellerSanchez2009}
{M{\"u}ller S{\'a}nchez}, F.~M., {Davies}, R.~I., {Genzel}, R., {Tacconi},
  L.~J., {Eisenhauer}, F., {Hicks}, E.~K.~S., {Friedrich}, S., \& {Sternberg},
  A. 2009, \apj, 691, 749

\bibitem[{{Nenkova} {et~al.}(2008){Nenkova}, {Sirocky}, {Nikutta},
  {Ivezi{\'c}}, \& {Elitzur}}]{Nenkova2008II}
{Nenkova}, M., {Sirocky}, M.~M., {Nikutta}, R., {Ivezi{\'c}}, {\v Z}., \&
  {Elitzur}, M. 2008, \apj, 685, 160

\bibitem[{{Neufeld} \& {Maloney}(1995)}]{Neufeld19954258}
{Neufeld}, D.~A., \& {Maloney}, P.~R. 1995, \apjl, 447, L17

\bibitem[{{Neufeld} {et~al.}(1994){Neufeld}, {Maloney}, \&
  {Conger}}]{Neufeld1994maser}
{Neufeld}, D.~A., {Maloney}, P.~R., \& {Conger}, S. 1994, \apjl, 436, L127

\bibitem[{{Nikola} {et~al.}(2011){Nikola}, {Stacey}, {Brisbin}, {Ferkinhoff},
  {Hailey-Dunsheath}, {Parshley}, \& {Tucker}}]{Nikola2011}
{Nikola}, T., {Stacey}, G.~J., {Brisbin}, D., {Ferkinhoff}, C.,
  {Hailey-Dunsheath}, S., {Parshley}, S., \& {Tucker}, C. 2011, \apj, 742, 88

\bibitem[{{Ogle} {et~al.}(2003){Ogle}, {Brookings}, {Canizares}, {Lee}, \&
  {Marshall}}]{Ogle2003}
{Ogle}, P.~M., {Brookings}, T., {Canizares}, C.~R., {Lee}, J.~C., \&
  {Marshall}, H.~L. 2003, \aap, 402, 849

\bibitem[{{Panuzzo} {et~al.}(2010){Panuzzo}, {Rangwala}, {Rykala}, {Isaak},
  {Glenn}, {Wilson}, {Auld}, {Baes}, {Barlow}, {Bendo}, {Bock}, {Boselli},
  {Bradford}, {Buat}, {Castro-Rodr{\'{\i}}guez}, {Chanial}, {Charlot},
  {Ciesla}, {Clements}, {Cooray}, {Cormier}, {Cortese}, {Davies}, {Dwek},
  {Eales}, {Elbaz}, {Fulton}, {Galametz}, {Galliano}, {Gear}, {Gomez},
  {Griffin}, {Hony}, {Levenson}, {Lu}, {Madden}, {O'Halloran}, {Okumura},
  {Oliver}, {Page}, {Papageorgiou}, {Parkin}, {P{\'e}rez-Fournon}, {Pohlen},
  {Polehampton}, {Rigby}, {Roussel}, {Sacchi}, {Sauvage}, {Schulz}, {Schirm},
  {Smith}, {Spinoglio}, {Stevens}, {Srinivasan}, {Symeonidis}, {Swinyard},
  {Trichas}, {Vaccari}, {Vigroux}, {Wozniak}, {Wright}, \&
  {Zeilinger}}]{Panuzzo2010}
{Panuzzo}, P. {et~al.} 2010, \aap, 518, L37

\bibitem[{{Papadopoulos} {et~al.}(2007){Papadopoulos}, {Isaak}, \& {van der
  Werf}}]{Papadopoulos2007}
{Papadopoulos}, P.~P., {Isaak}, K.~G., \& {van der Werf}, P.~P. 2007, \apj,
  668, 815

\bibitem[{{Papadopoulos} \&
  {Seaquist}(1999{\natexlab{a}})}]{Papadopoulos1999SCUBA}
{Papadopoulos}, P.~P., \& {Seaquist}, E.~R. 1999{\natexlab{a}}, \apjl, 514, L95

\bibitem[{{Papadopoulos} \&
  {Seaquist}(1999{\natexlab{b}})}]{Papadopoulos1999CO}
---. 1999{\natexlab{b}}, \apj, 516, 114

\bibitem[{{Pier} {et~al.}(1994){Pier}, {Antonucci}, {Hurt}, {Kriss}, \&
  {Krolik}}]{Pier1994}
{Pier}, E.~A., {Antonucci}, R., {Hurt}, T., {Kriss}, G., \& {Krolik}, J. 1994,
  \apj, 428, 124

\bibitem[{{Pilbratt} {et~al.}(2010){Pilbratt}, {Riedinger}, {Passvogel},
  {Crone}, {Doyle}, {Gageur}, {Heras}, {Jewell}, {Metcalfe}, {Ott}, \&
  {Schmidt}}]{Pilbratt2010}
{Pilbratt}, G.~L. {et~al.} 2010, \aap, 518, L1+

\bibitem[{{Planesas} {et~al.}(1991){Planesas}, {Scoville}, \&
  {Myers}}]{Planesas1991}
{Planesas}, P., {Scoville}, N., \& {Myers}, S.~T. 1991, \apj, 369, 364

\bibitem[{{Poelman} \& {Spaans}(2005)}]{Poelman2005}
{Poelman}, D.~R., \& {Spaans}, M. 2005, \aap, 440, 559

\bibitem[{{Poglitsch} {et~al.}(2010){Poglitsch}, {Waelkens}, {Geis},
  {Feuchtgruber}, {Vandenbussche}, {Rodriguez}, {Krause}, {Renotte}, {van
  Hoof}, {Saraceno}, {Cepa}, {Kerschbaum}, {Agn{\`e}se}, {Ali}, {Altieri},
  {Andreani}, {Augueres}, {Balog}, {Barl}, {Bauer}, {Belbachir}, {Benedettini},
  {Billot}, {Boulade}, {Bischof}, {Blommaert}, {Callut}, {Cara}, {Cerulli},
  {Cesarsky}, {Contursi}, {Creten}, {De Meester}, {Doublier}, {Doumayrou},
  {Duband}, {Exter}, {Genzel}, {Gillis}, {Gr{\"o}zinger}, {Henning},
  {Herreros}, {Huygen}, {Inguscio}, {Jakob}, {Jamar}, {Jean}, {de Jong},
  {Katterloher}, {Kiss}, {Klaas}, {Lemke}, {Lutz}, {Madden}, {Marquet},
  {Martignac}, {Mazy}, {Merken}, {Montfort}, {Morbidelli}, {M{\"u}ller},
  {Nielbock}, {Okumura}, {Orfei}, {Ottensamer}, {Pezzuto}, {Popesso},
  {Putzeys}, {Regibo}, {Reveret}, {Royer}, {Sauvage}, {Schreiber}, {Stegmaier},
  {Schmitt}, {Schubert}, {Sturm}, {Thiel}, {Tofani}, {Vavrek}, {Wetzstein},
  {Wieprecht}, \& {Wiezorrek}}]{Poglitsch2010}
{Poglitsch}, A. {et~al.} 2010, \aap, 518, L2+

\bibitem[{{Raban} {et~al.}(2009){Raban}, {Jaffe}, {R{\"o}ttgering},
  {Meisenheimer}, \& {Tristram}}]{Raban2009}
{Raban}, D., {Jaffe}, W., {R{\"o}ttgering}, H., {Meisenheimer}, K., \&
  {Tristram}, K.~R.~W. 2009, \mnras, 394, 1325

\bibitem[{{Rangwala} {et~al.}(2011){Rangwala}, {Maloney}, {Glenn}, {Wilson},
  {Rykala}, {Isaak}, {Baes}, {Bendo}, {Boselli}, {Bradford}, {Clements},
  {Cooray}, {Fulton}, {Imhof}, {Kamenetzky}, {Madden}, {Mentuch}, {Sacchi},
  {Sauvage}, {Schirm}, {Smith}, {Spinoglio}, \& {Wolfire}}]{Rangwala2011}
{Rangwala}, N. {et~al.} 2011, \apj, 743, 94

\bibitem[{{Rigopoulou} {et~al.}(2002){Rigopoulou}, {Kunze}, {Lutz}, {Genzel},
  \& {Moorwood}}]{Rigopoulou2002}
{Rigopoulou}, D., {Kunze}, D., {Lutz}, D., {Genzel}, R., \& {Moorwood},
  A.~F.~M. 2002, \aap, 389, 374

\bibitem[{{Rodr{\'{\i}}guez-Ardila} {et~al.}(2005){Rodr{\'{\i}}guez-Ardila},
  {Riffel}, \& {Pastoriza}}]{Rodriguez_Ardila2005}
{Rodr{\'{\i}}guez-Ardila}, A., {Riffel}, R., \& {Pastoriza}, M.~G. 2005,
  \mnras, 364, 1041

\bibitem[{{Rotaciuc} {et~al.}(1991){Rotaciuc}, {Krabbe}, {Cameron}, {Drapatz},
  {Genzel}, {Sternberg}, \& {Storey}}]{Rotaciuc1991}
{Rotaciuc}, V., {Krabbe}, A., {Cameron}, M., {Drapatz}, S., {Genzel}, R.,
  {Sternberg}, A., \& {Storey}, J.~W.~V. 1991, \apjl, 370, L23

\bibitem[{{Roussel} {et~al.}(2007){Roussel}, {Helou}, {Hollenbach}, {Draine},
  {Smith}, {Armus}, {Schinnerer}, {Walter}, {Engelbracht}, {Thornley},
  {Kennicutt}, {Calzetti}, {Dale}, {Murphy}, \& {Bot}}]{Roussel2007}
{Roussel}, H. {et~al.} 2007, \apj, 669, 959

\bibitem[{{Schartmann} {et~al.}(2010){Schartmann}, {Burkert}, {Krause},
  {Camenzind}, {Meisenheimer}, \& {Davies}}]{Schartmann2010}
{Schartmann}, M., {Burkert}, A., {Krause}, M., {Camenzind}, M., {Meisenheimer},
  K., \& {Davies}, R.~I. 2010, \mnras, 403, 1801

\bibitem[{{Schinnerer} {et~al.}(2000){Schinnerer}, {Eckart}, {Tacconi},
  {Genzel}, \& {Downes}}]{Schinnerer2000}
{Schinnerer}, E., {Eckart}, A., {Tacconi}, L.~J., {Genzel}, R., \& {Downes}, D.
  2000, \apj, 533, 850

\bibitem[{{Schleicher} {et~al.}(2010){Schleicher}, {Spaans}, \&
  {Klessen}}]{Schleicher2010alma}
{Schleicher}, D.~R.~G., {Spaans}, M., \& {Klessen}, R.~S. 2010, \aap, 513, A7+

\bibitem[{{Sempere} {et~al.}(2000){Sempere}, {Cernicharo}, {Lefloch},
  {Gonz{\'a}lez-Alfonso}, \& {Leeks}}]{Sempere2000}
{Sempere}, M.~J., {Cernicharo}, J., {Lefloch}, B., {Gonz{\'a}lez-Alfonso}, E.,
  \& {Leeks}, S. 2000, \apjl, 530, L123

\bibitem[{{Spaans} \& {Meijerink}(2008)}]{Spaans2008}
{Spaans}, M., \& {Meijerink}, R. 2008, \apjl, 678, L5

\bibitem[{{Spinoglio} {et~al.}(2005){Spinoglio}, {Malkan}, {Smith},
  {Gonz{\'a}lez-Alfonso}, \& {Fischer}}]{Spinoglio2005}
{Spinoglio}, L., {Malkan}, M.~A., {Smith}, H.~A., {Gonz{\'a}lez-Alfonso}, E.,
  \& {Fischer}, J. 2005, \apj, 623, 123

\bibitem[{{Sternberg} \& {Dalgarno}(1989)}]{Sternberg1989}
{Sternberg}, A., \& {Dalgarno}, A. 1989, \apj, 338, 197

\bibitem[{{Sternberg} \& {Dalgarno}(1995)}]{Sternberg1995}
---. 1995, \apjs, 99, 565

\bibitem[{{Sternberg} {et~al.}(1994){Sternberg}, {Genzel}, \&
  {Tacconi}}]{Sternberg1994}
{Sternberg}, A., {Genzel}, R., \& {Tacconi}, L. 1994, \apjl, 436, L131

\bibitem[{{Sturm} {et~al.}(2002){Sturm}, {Lutz}, {Verma}, {Netzer},
  {Sternberg}, {Moorwood}, {Oliva}, \& {Genzel}}]{Sturm2002}
{Sturm}, E., {Lutz}, D., {Verma}, A., {Netzer}, H., {Sternberg}, A.,
  {Moorwood}, A.~F.~M., {Oliva}, E., \& {Genzel}, R. 2002, \aap, 393, 821

\bibitem[{{Tacconi} {et~al.}(1994){Tacconi}, {Genzel}, {Blietz}, {Cameron},
  {Harris}, \& {Madden}}]{Tacconi1994}
{Tacconi}, L.~J., {Genzel}, R., {Blietz}, M., {Cameron}, M., {Harris}, A.~I.,
  \& {Madden}, S. 1994, \apjl, 426, L77+

\bibitem[{{Telesco} \& {Harper}(1980)}]{Telesco1980}
{Telesco}, C.~M., \& {Harper}, D.~A. 1980, \apj, 235, 392

\bibitem[{{Thompson} {et~al.}(1978){Thompson}, {Lebofsky}, \&
  {Rieke}}]{Thompson1978}
{Thompson}, R.~I., {Lebofsky}, M.~J., \& {Rieke}, G.~H. 1978, \apjl, 222, L49

\bibitem[{{Tomono} {et~al.}(2006){Tomono}, {Terada}, \&
  {Kobayashi}}]{Tomono2006}
{Tomono}, D., {Terada}, H., \& {Kobayashi}, N. 2006, \apj, 646, 774

\bibitem[{{Usero} {et~al.}(2004){Usero}, {Garc{\'{\i}}a-Burillo}, {Fuente},
  {Mart{\'{\i}}n-Pintado}, \& {Rodr{\'{\i}}guez-Fern{\'a}ndez}}]{Usero2004}
{Usero}, A., {Garc{\'{\i}}a-Burillo}, S., {Fuente}, A.,
  {Mart{\'{\i}}n-Pintado}, J., \& {Rodr{\'{\i}}guez-Fern{\'a}ndez}, N.~J. 2004,
  \aap, 419, 897

\bibitem[{{van der Werf} {et~al.}(2010){van der Werf}, {Isaak}, {Meijerink},
  {Spaans}, {Rykala}, {Fulton}, {Loenen}, {Walter}, {Wei{\ss}}, {Armus},
  {Fischer}, {Israel}, {Harris}, {Veilleux}, {Henkel}, {Savini}, {Lord},
  {Smith}, {Gonz{\'a}lez-Alfonso}, {Naylor}, {Aalto}, {Charmandaris}, {Dasyra},
  {Evans}, {Gao}, {Greve}, {G{\"u}sten}, {Kramer}, {Mart{\'{\i}}n-Pintado},
  {Mazzarella}, {Papadopoulos}, {Sanders}, {Spinoglio}, {Stacey}, {Vlahakis},
  {Wiedner}, \& {Xilouris}}]{vanderWerf2010}
{van der Werf}, P.~P. {et~al.} 2010, \aap, 518, L42+

\bibitem[{{van Kempen} {et~al.}(2010){van Kempen}, {Kristensen}, {Herczeg},
  {Visser}, {van Dishoeck}, {Wampfler}, {Bruderer}, {Benz}, {Doty}, {Brinch},
  {Hogerheijde}, {J{\o}rgensen}, {Tafalla}, {Neufeld}, {Bachiller}, {Baudry},
  {Benedettini}, {Bergin}, {Bjerkeli}, {Blake}, {Bontemps}, {Braine},
  {Caselli}, {Cernicharo}, {Codella}, {Daniel}, {di Giorgio}, {Dominik},
  {Encrenaz}, {Fich}, {Fuente}, {Giannini}, {Goicoechea}, {de Graauw},
  {Helmich}, {Herpin}, {Jacq}, {Johnstone}, {Kaufman}, {Larsson}, {Lis},
  {Liseau}, {Marseille}, {McCoey}, {Melnick}, {Nisini}, {Olberg}, {Parise},
  {Pearson}, {Plume}, {Risacher}, {Santiago-Garc{\'{\i}}a}, {Saraceno},
  {Shipman}, {van der Tak}, {Wyrowski}, {Y{\i}ld{\i}z}, {Ciechanowicz},
  {Dubbeldam}, {Glenz}, {Huisman}, {Lin}, {Morris}, {Murphy}, \&
  {Trappe}}]{vanKempen2010CO}
{van Kempen}, T.~A. {et~al.} 2010, \aap, 518, L121

\bibitem[{{Wada} {et~al.}(2009){Wada}, {Papadopoulos}, \&
  {Spaans}}]{Wada2009AGN}
{Wada}, K., {Papadopoulos}, P.~P., \& {Spaans}, M. 2009, \apj, 702, 63

\bibitem[{{Ward} {et~al.}(2003){Ward}, {Zmuidzinas}, {Harris}, \&
  {Isaak}}]{Ward2003}
{Ward}, J.~S., {Zmuidzinas}, J., {Harris}, A.~I., \& {Isaak}, K.~G. 2003, \apj,
  587, 171

\bibitem[{{Wilson} \& {Ulvestad}(1987)}]{Wilson1987}
{Wilson}, A.~S., \& {Ulvestad}, J.~S. 1987, \apj, 319, 105

\bibitem[{{Yang} {et~al.}(2010){Yang}, {Stancil}, {Balakrishnan}, \&
  {Forrey}}]{Yang2010}
{Yang}, B., {Stancil}, P.~C., {Balakrishnan}, N., \& {Forrey}, R.~C. 2010,
  \apj, 718, 1062

\end{thebibliography}
\end{document}